\documentstyle[prd,aps,floats,psfig]{revtex}
\def\spose#1{\hbox to 0pt{#1\hss}}
\def\lta{\mathrel{\spose{\lower 3pt\hbox{$\mathchar"218$}}
     \raise 2.0pt\hbox{$\mathchar"13C$}}}
\tighten
\begin{document} 
\draft
\preprint{
\begin{tabular}{rr}
CfPA/96-th-15\\
\end{tabular}
}

\title{Cosmology with a Primordial Scaling Field}
\author{Pedro G. Ferreira$^{1}$ and Michael Joyce$^{2}$ }
\address{$^{1}$Center for Particle Astrophysics, 301 Leconte Hall,
 University of California,
Berkeley, CA 94720\\
$^{2}$School of Mathematics, Trinity College, Dublin 2, Ireland}
\maketitle
\begin{abstract}
A weakly coupled scalar field $\Phi$ with a simple exponential potential 
$V=M_P^4\exp(-\lambda\Phi/M_P)$ where $M_P$
is the reduced Planck mass, and $\lambda > 2$, has an 
attractor solution in a radiation or matter dominated universe
in which it mimics the scaling of the dominant component, 
contributing a fixed fraction $\Omega_\phi$ (determined by $\lambda$) to
the energy density. Such fields arise generically in particle physics 
theories involving compactified dimensions, with values of
$\lambda$ which give a cosmologically relevant $\Omega_\phi$.
 For natural initial conditions on the scalar field in 
the early universe the attractor solution is established long before 
the epoch of structure formation, and in contrast to the solutions
used in other scalar field cosmologies, it is one which
does not involve an energy scale for the scalar field 
characteristic of late times .
We study in some detail the evolution of matter and radiation perturbations 
in a standard inflation-motivated $\Omega=1$ dark-matter dominated
cosmology with this extra field. Using a full 
Einstein-Boltzmann calculation we compare
observable quantities with current data. We find
that, for $\Omega_\phi\simeq 0.08-0.12$, these models
are consistent with  large angle cosmic microwave background
anisotropies as detected by COBE, the linear mass variance
as compiled from galaxy surveys, big bang nucleosynthesis, the
abundance of rich clusters and constraints from the Lyman-$\alpha$ 
systems at high redshift.
Given the simplicity of the model, its theoretical motivation 
and its success in matching observations, we argue that it
should be taken on a par with other currently viable 
models of structure formation\end{abstract}

\date{\today}


\renewcommand{\thefootnote}{\arabic{footnote}}
\setcounter{footnote}{0}
\section{Introduction}
The past twenty years have seen a tremendous revolution in how
we study the origin and evolution of our universe. On the one hand
developments in theoretical particle physics have lead to a proliferation
of ideas on how high energy physics might have an observable effect
on the large scale structure of our universe. On the other hand
the increasing quality of astrophysical data has led to firm
constraints on what physics is allowed in the early universe. 
Probably  the most impressive example of such an interplay is how the COBE
detection \cite{smoot} has affected the most popular and theoretically
explored theory of structure formation, the standard cold dark 
matter model (SCDM). 

The SCDM model brings together the idea of inflation
 \cite{guth,linde,as}
 and the picture of large scale gravitational collapse
\cite{peebles}. A period of superluminal
expansion of the universe 
would have led to the amplification of subhorizon vacuum fluctuations
to superhorizon scales. The net result would be a set of scale
invariant, Gaussian perturbations which would evolve, through
gravitational collapse, into the structures we see today. The relic
radiation bears an imprint of these fluctuations from when the
universe recombined.  The distribution of galaxies and clusters of
galaxies should reflect these fluctuations today.
It has been found however that the SCDM model cannot successfully
accommodate the data observed on all scales.
Matching its predictions to COBE measurements (on large scales) 
of the microwave
background, one finds that the amplitude of fluctuations
on $8$h$^{-1}$Mpc scales (where h is the Hubble constant today in
units of $100$km/s/Mpc ) do not match those observed\cite{dod}. 
To be more specific, if we define the amplitude of mass fluctuations
in a sphere of radius $8$h$^{-1}$Mpc, $\sigma_8$, then COBE normalized
SCDM predicts $\sigma_8=1.2$ while the measured value (through
cluster abundances) is $\sigma_8=0.6\pm0.1$, a discrepancy by a factor
of 2.   Further not only the amplitude but also the {\it scale dependence}
of the SCDM model differs from the one measured \cite{PD},  and there
are also a number of problems with the non-linear evolution of baryons and
velocities on small scales. 

These failings of SCDM have led to attempts to modify it,
while keeping its basic features intact. The latter 
(a simple choice of background cosmological parameters
with $\Omega=1$ and quantum generation of fluctuations) are 
associated with its theoretical motivation from inflation.
The most prominent candidate theories of structure formation 
of this type are now a universe with a cosmological constant ($\Lambda$CDM)
\cite{lambda-models}
and a universe with a fifth of its dark matter component
in two families of massive neutrinos (MDM)\cite{mdm-models}. 
Models of structure formation in an open universe (OCDM), for
which there is considerable evidence, have also been extensively
studied \cite{open-models}. All these models, once COBE normalized, 
predict the approximately correct distribution of mass fluctuations. 
Over the past few years other flat models have been constructed which, 
like MDM, are more related to our understanding of fundamental 
particle physics. This is the case of decaying cosmological constant 
models \cite{fhsw,cdf} and decaying massive particles \cite{dod,gws}. 
Unfortunately, unlike the SCDM scenario, all these models involve
a tuning of parameters which is unnatural from the point of view
of particle physics, simply because one is using super-GeV physics 
to explain sub-eV observations. Just as is the case of 
$\Lambda$CDM (which involves tuning the cosmological constant
to be relevant only at present epochs) this tuning does not provide
a reason to discard these models, but is a very unattractive feature 
of them.

Both the cosmology of weakly coupled scalar fields and their 
theoretical motivation have been much studied since the 
advent of the idea of inflation.
In most of the theoretical particle physics extensions of 
the standard model (e.g. supersymmetry and supergravity) 
weakly coupled scalar fields of various kinds are ubiquitous. 
Unfortunately there is as yet no experimental evidence
for the existence of any fundamental scalar field at all, 
in particular the Higgs particle of the standard model
remains to date undetected. However the theoretical motivation 
for such scalar fields is sufficiently compelling 
that it is certainly worthwhile considering what consequences  
their existence might have for cosmology beyond the confines of
inflation. An example of this is given by the main alternative 
of structure formation - defect theories - which usually
involve the existence of some scalar field e.g. the `texture' 
theory \cite{turok} relies on the existence of a scalar field invariant 
under a global non-abelian symmetry broken at the GUT scale,
with associated goldstone particles which are unobservably weakly
coupled to ordinary matter.

The role that a weakly coupled scalar field might play in late
time cosmology if it contributes a component to the homogeneous
background energy density of the universe has also been 
investigated. In \cite{RP} the authors  considered in general
terms the idea that a significant contribution to the energy density
from a homogeneous scalar field could have an important effect
on structure formation, and applied the idea to a baryonic
universe with a initial spectrum of perturbations with power law 
behaviour. After some general analysis of the homogeneous dynamics,
the model singled out for a detailed treatment was that of a
scalar field in a negative power law potential which comes to 
dominate at late times, producing behaviour very similar to that 
associated with a cosmological constant . Another kind of model which 
has been developed in detail, for the case of a cold
dark matter dominated cosmology, in \cite{fhsw,cdf} involves a 
cosine potential with a combination of Planck and sub $eV$ physics.
This provides a specific realisation of a 
`decaying cosmological constant', in which the field
initially behaves like a cosmological constant and
then rolls so that it scales asymptotically like matter. 
All these models have an energy density in the scalar field
which comes to dominate at late times, as have two more recent
very detailed studies \cite{steinhardt,liddleviana}.  
In \cite{steinhardt} the case of a scalar field which scales slower
than matter is  described more generically in terms of its equation 
of state, and in \cite{liddleviana} the specific cases of late
time dominance realized in both cosine and exponential potentials.

In this paper we present in detail the results which have
been reported in \cite{FJ}. The model we study is
SCDM with the addition of a scalar field $\Phi$ with a
simple exponential potential $V=M_P^4\exp(-\lambda\Phi/M_P)$ 
where $M_P\equiv(8\pi G)^{-\frac{1}{2}}=2.4 \times 10^{18}$GeV
is the reduced Planck mass, and $\lambda > 2$. In this specific
case there is a very special and interesting solution for 
matter and radiation coupled through gravity to this field,
in which the scalar field energy follows that of the dominant 
component, contributing a fixed fraction of the total energy
density determined by $\lambda$. The existence of this
homogeneous solution was shown in \cite{RP,wett,clw}.
Because the scalar field can contribute at most a small fraction of the 
total energy density at nucleosynthesis, its potential
interest in the context of the problem of structure formation  
has been overlooked. This constraint suggests {\it prima facie}
that the field in such an attractor can have little effect on
cosmology at late time. As we shall see, this is incorrect,
for the simple reason that a small contribution acting over
a long time can have as big an effect as a large contribution 
entering only at late times. The particular merit of the model 
is related to the fact that the cosmological solution is an attractor for
the scalar field: Because of this, there is no tuning
of the type involved in {\it all} other proposed
modifications of SCDM. The only parameter additional to SCDM 
is $\lambda$, and the value $(\sim 5-6)$ which gives a best 
fit to structure formation, is of the order naturally expected in the
particle physics models in which the potential arises. 
As a cosmology it resembles MDM much more than any of the
scalar field models which have been studied in the literature.

Some comment is perhaps necessary at the outset regarding
the assumption that the universe is flat and dominated by
a component scaling as matter, as there is mounting 
observational evidence that this is not the case. 
The most advertised is the age problem, that is, the fact that 
we see objects which are older than the age of the universe if 
we assume a flat matter dominated universe with the currently 
observed Hubble constant of $H_0=65\pm 10$kms$^{-1}$Mpc$^{-1}$. 
Several of the modifications we have discussed (in particular
the cosmological constant model and some of the scalar field models)
avoid this problem since the dominant component at the present epoch
is not matter. The recent measurements of the Hipparcos satellite 
\cite{hipparcos}
seem to indicate that the age of these objects have been overestimated 
by around 10$\%$ and a re-analysis of the uncertainties in the estimates 
seem to indicate that a flat matter dominated universe with 
$H_0<66$kms$^{-1}$Mpc$^{-1}$ is compatible with the current age estimates. 
Another argument for a low density universe
comes from the analysis of large scale flows. In \cite{davis} it
has been argued that $\beta=\Omega^{0.6}/b<1$ from the analysis of
the MARK III data and comparison with the IRAS surveys. However
there is concern with the self consistency of the velocity data
and another group \cite{dekel} indicates that the the flows are
consistent with a universe with $\Omega=0.4-1$. Small scale
observations also seem to indicate that there are problems with a
high density universe. In particular the baryon fraction in clusters is
difficult to reconcile with the BBN limits unless one considers
a low density universe \cite{Evrard}. Again there are uncertainties in such
an analysis; they rely on elaborate numerical simulations which
are at the limit of current computational power. It is conceivable
that some physics (such as cooling of the cluster medium) is
being overlooked. The same can be said for the cold velocity
dispersion which is measured on few Mpc scales \cite{peebles}.
Although all these observations together begin to make a strong
case for an open universe, the uncertainties and inconsistencies
are sufficiently large for us to still consider a high density, matter
dominated universe.
It may be that in the next few years the evidence is sufficiently
compelling to rule out these models, but this is definitely
not the case yet.

The structure of the paper is as follows. In section \ref{PP} 
we first discuss the homogeneous modes of scalar fields in a 
universe with matter and radiation in general terms, and explain
how the case of a simple exponential with its attractor solutions
for $\lambda >2$ is a special one. We discuss briefly the possible origin of 
exponential potentials in fundamental theories. We then discuss 
the initial conditions on this scalar field in the early universe,
and when these lead the attractor to be established. In typical 
inflationary theories we argue that the attractor 
describes the homogeneous cosmology well before nucleosynthesis,
while in an alternative theory of reheating which can be realised with
the same exponential field the scalar field may still be negligible
at nucleosynthesis for natural parameter values.
In section \ref{pert} we analyse the evolution of perturbations in 
our scenario. We describe the complete set of equations which 
govern their evolution and analyse the asymptotic behaviour in 
the interesting regimes. The similarities
with evolution of density fluctuations in an MDM universe lead us
to pursue the comparison in some detail and we come to the
conclusion that scalar field is more effective at suppressing
perturbations than hot dark matter. We also consider the effect
on the cosmic microwave background (CMB) and deconstruct the different
effects it has on the angular power spectrum of the CMB. In section 
\ref{data} we compare the predictions of linear perturbation theory
(using the full Boltzmann-Einstein equations) with some observations.
In particular we compare it to the COBE data and then use this
comparison to normalize the theory and compare to the Peacock and
Dodds data. As a byproduct we derive a fit to $\sigma_8$ for our
models and we quantify the amount of structure on small
scales and high redshift, comparing with constraints derived from 
Lyman-$\alpha$ systems.
 In \ref{conc} we summarise our findings and draw some conclusions
about future prospects for our model and other issues related
to it which might be further investigated.

\section{The `Self-Tuning'  Scalar Field with an Exponential Potential}
\label{PP}

In this section we describe in detail the homogeneous attractor 
solutions which specify the zero-order cosmology, about which we
treat perturbations fully in the subsequent part of the paper.
We explain the very special feature of this model which contrasts
it with other scalar field cosmologies which have been treated
in the literature: {\it No special energy scale characteristic of
late time cosmology need be invoked to produce the required 
solution}. We will show that for a very wide and
natural range of initial conditions on the scalar field in the very
early universe, the attractor will be attained as early as assumed
in the rest of the paper. 

\subsection{Scaling of the Energy Density in a Scalar Field}

We begin with a general discussion of homogeneous 
FRW cosmology in which there is, in addition to the usual 
matter and radiation content, a contribution to the energy 
momentum coming from a scalar field $\Phi$ with a potential
$V(\Phi)$. In the rest of the paper we use the notation 
$\Phi=\phi+\varphi$, where $\varphi$ denotes the perturbation
about the homogeneous solution $\phi$.   
Working for the moment in comoving
coordinates in which the metric is $ds^2=-dt^2 + a^2 \delta_{ij}dx^i dx^j$
(where $a$ is the scale factor),
the contribution to the energy-momentum tensor from the scalar field is 
\begin{equation}
T_{00}= \rho_\phi \qquad T_{ij}=a^2 p_\phi\delta_{ij}  \qquad
{\rm where} \qquad 
\rho_{\phi}= \left [{1 \over 2 }{\dot \phi}^2+V(\phi) \right ] \qquad
p_{\phi}=\left [{1 \over 2 }{\dot \phi}^2- V(\phi)\right].
\label{background1}
\end{equation}
The equations of motion for the scalar field are then 
\begin{eqnarray}
\ddot{\phi} + 3 H \dot{\phi} +
V'(\phi)=\frac{1}{a^3}\frac{d}{dt}(a^3\dot{\phi}) + V'(\phi)=0
\label{eq: potleoma}\\
{H}^2 = \frac{1}{3M_{p}^2}(\frac{1}{2}\dot{\phi}^2 + V(\phi)+\rho_n) 
\label{eq: potleomb}\\
\dot{\rho_n} + n{H} \rho_n = 0
\label{eq: potleomc}
\end{eqnarray}
where $\rho_n$ is the energy density in radiation ($n=4$)
or non-relativistic matter ($n=3$), ${H}=\frac{\dot{a}}{a}$ 
is the Hubble expansion rate of the universe, 
dots are derivatives with respect to time, 
primes derivatives with respect to the field $\phi$,
and $M_P\equiv(8 \pi G)^{-\frac{1}{2}}=2.4 \times 10^{18}$GeV 
is the reduced Planck mass. The scalar field is assumed
to be coupled to ordinary matter only through gravity.
Multiplying (\ref{eq: potleoma}) by $\dot{\phi}$ and
integrating, one obtains     
\begin{equation}
\rho_\phi(a) 
= \rho(a_o) \exp \big(-\int_{a_o}^{a} 6(1 - \xi(a)) \frac 
{da}{a} \big) \qquad 
{\rm where} \qquad \rho_{\phi}=\frac{1}{2} \dot{\phi}^2+ V(\phi)
\qquad \xi= \frac{V(\phi)}{\rho_\phi}.
\label{eq: scaling}
\end{equation} 
It follows from this that, given the range of possible values
$0<\xi<1$ (assuming $V(\phi)$ is positive), the energy density 
of a scalar field has the range of scaling behaviours 
\begin{equation}
\rho_\phi  \propto 1/a^m \qquad 0 \leq m \leq 6
\label{possible-scaling}
\end{equation}
How the energy in a homogeneous mode of a scalar 
field scales is thus determined by the ratio of the
potential to the kinetic energy. Alternatively one can 
phrase the statement in terms of the equation of state obeyed 
by the mode: From (\ref{background1}) we have $\xi=\frac{1}{2}(1-w)$ where
$p_\phi=w \rho_\phi$, and $m=3(1+w)$ (for constant $w$)
in (\ref{eq: scaling}).

These statements are true independent of any specific assumption 
about ${H}$ i.e. about what dominates the 
energy density of the universe.  When the potential energy 
dominates over kinetic energy, we have $\xi \rightarrow 1$
and therefore $\rho_\phi \approx constant $ i.e. an energy density
which behaves like a cosmological constant (and $w=-1$); 
in the opposite limit of $\xi \rightarrow 0$ i.e. a
kinetic energy dominated mode, we have an energy density 
with $\rho_\phi \propto 1/a^6$ i.e. red-shifting faster
than radiation or matter (and $w=+1$). Inflation occurs 
when the former type of mode also dominates the energy density;
the opposite limit, when the universe is dominated by
the kinetic energy of a scalar field (which,
following \cite{mj}, we refer to as {\it kination})
gives a sub-luminal expansion with $a \propto t^{\frac{1}{3}}$.

We now consider more specifically what sorts of potentials 
give rise to these different types of scaling. One simple case 
to analyse is that in which a field rolls about the
minimum of a potential. In such an oscillatory
phase the approximate scaling of the energy density 
can be extracted from (\ref{eq: scaling}) by replacing
$\xi $ by its average value over an oscillation. For a potential
which is power law about its minimum with $V(\phi) \propto \phi^n$ 
the result \cite{turner} is that 
\begin{equation}
\xi=\frac{2}{n+2} \qquad \rho_\phi \propto 1/a^m \qquad m=\frac{6n}{n+2}
\label{scaling-oscillation}
\end{equation}
reproducing the well known result that a coherent mode
oscillating in a quadratic potential gives the same scaling as
matter, and a $\phi^4$ potential that of radiation. For $n>4$
one obtains modes scaling faster than radiation. 

Again this statement does not depend on what component dominates 
the energy density, and the same scaling applies to the mode 
irrespective of whether the universe has 
$\rho_n=0$, or is matter or radiation dominated. 
The case of a field rolling down a potential (before it
reaches its minimum, or if it has no minimum) is quite
different. The equation of motion (\ref{eq: potleoma}) is just 
a damped roll with the energy content determining the damping 
through (\ref{eq: potleomb}). The scaling obtained for a given potential  
depends on what components are present, because what determines 
the scaling is the balance between the increase in kinetic energy relative
to potential energy as the field rolls down the potential, and 
the decrease of the same quantity due to the damping.
The criterion for a particular scaling is therefore a
requirement of the `steepness' of the potential. 
That the simple exponential potential $V(\phi)=V_o e^{-\lambda\phi /M_P }$
provides the appropriate yard-stick in the case of scalar field
domination is indicated by the existence of a family of 
solutions for this potential to 
(\ref{eq: potleoma})-(\ref{eq: potleomb}) with
$\rho_n=0$ \cite{RP,wett,clw} 
\begin{equation}
\phi(t) = \phi_o+ \frac{2M_P}{\lambda} \ln (t M_P) 
\qquad \phi_o= \frac{2M_P}{\lambda} \ln(\frac {V_o\lambda^2 }{2M_P^4(6-\lambda^2)} )
\qquad \xi=1-\frac{\lambda^2}{6}  
\qquad \rho_\phi \propto \frac{1}{a^{\lambda^2}}
\qquad a \propto t^{2/\lambda^2}
\label{eq: jhsoln}
\end{equation}
for $\lambda < \sqrt{6}$, and $\phi_o$ can always be chosen
to be zero by redefining the origin of $\phi$, in which case
$V_o=\frac{2}{\lambda^2}(\frac{6}{\lambda^2}-1)M_P^4$.
These solutions, which are attractors\footnote{It is 
simple to verify directly from (\ref{eq: potleoma})
and (\ref{eq: potleomb}) that homogeneous perturbations 
about the solution decay as $t^{-1}$ and $t^{1- \frac{6}{\lambda^2}}$.}  
were written down \cite{halliwellattractors,barrowattractors} in 
the context of power-law inflation \cite{power-law-inflation}, 
associated with the superluminal
growth of the scale-factor for $\lambda < \sqrt {2}$.  
For $\lambda > 6$ there is not a single attractor with
a finite value of $\xi$, but every solution has 
$\xi \rightarrow 0$ asymptotically and $\rho \propto 1/a^6$. 

As $\lambda$ increases from zero we obtain the entire range of 
possible scaling behaviours (\ref{possible-scaling})
for the energy in a scalar field. By comparison with this potential
we can infer how scalar modes will  scale: The `slow-roll' conditions 
($|M_P V'/V| << \sqrt{6} $, $|M_P^2 V''/V| << 3 $) for inflation,
for example, could be stated as the requirement that the first
two derivatives of a potential be smaller than those
of an exponential with $\lambda= \sqrt{3}$. An analagous 
`fast-roll' condition for `kination' can clearly be provided
by comparison with an exponential with $\lambda= \sqrt 6$
e.g. a potential $\sim e^{-\mu \phi^2/M_P^2}$ will have
such modes for sufficiently large $\phi$.

\begin{figure}
\centerline{\psfig{file=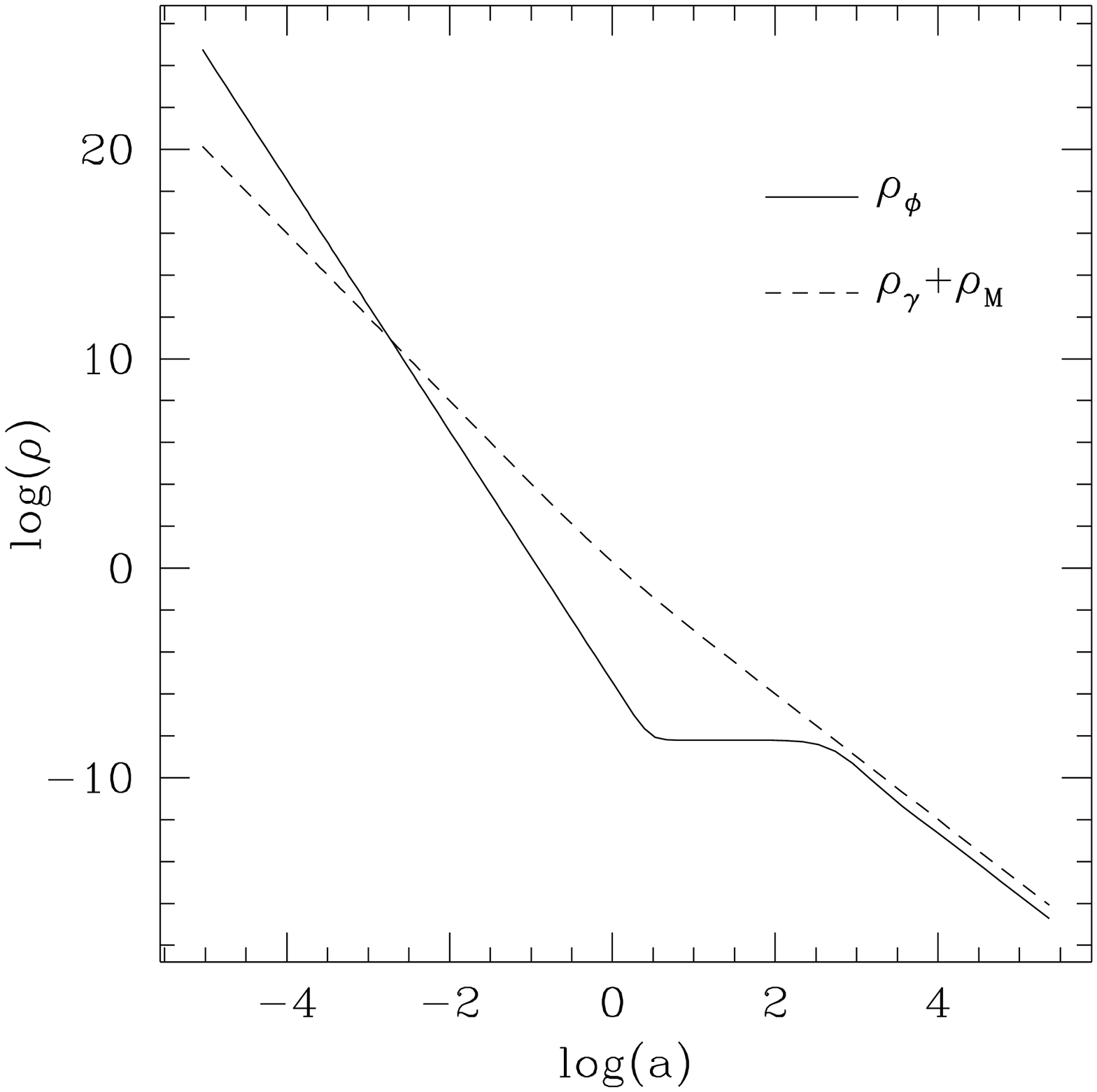,width=3.5in}
\psfig{file=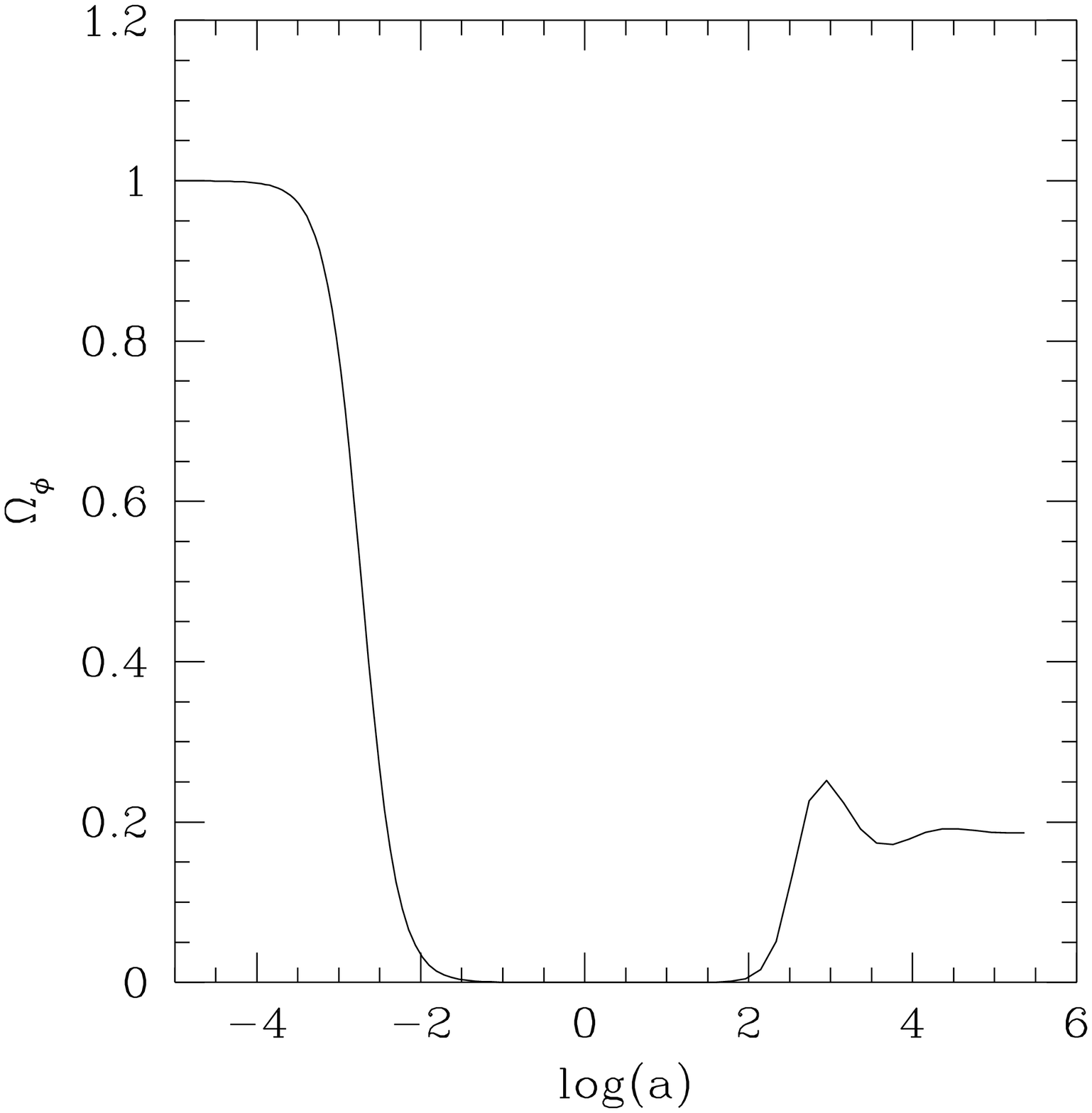,width=3.5in}}
\caption{In the left panel we plot the evolution of the energy density
in the scalar field ($\rho_\phi$) and in a component of radiation-matter 
as a function of scale factor for a situation in which the scalar field
(with $\lambda=4$) initially dominates, then undergoes a transient and finally locks on to the
scaling solution. In the right panel we plot the evolution of the
fractional density in the scalar field.}\label{transient}
\end{figure}

These statements apply only to the case of scalar
field dominance since we took $\rho_n=0$.  What interests us
in the present context is how the scalar energy behaves in
the presence of matter and radiation i.e. with $\rho_n \neq 0$. 
There are two quite distinct cases which can be immediately distinguished
according to their behaviour when $\rho_n=0$:
Those potentials in which the energy density scales
slower than $1/a^n$ (i.e. $\lambda < \sqrt{n}$), and
those in which it scales faster ($\lambda > \sqrt{n}$). 
When $\rho_n \neq 0$ these two types of potential will show 
very different behaviour for the following reason:  Adding 
a component increases the damping term in (\ref{eq: potleoma}), 
and it follows that the scaling with $a$ of the energy density 
in the scalar field is always {\it slower} 
i.e. $\rho_\phi \propto 1/a^{\lambda^2 - \delta}$
with $\lambda^2 \geq \delta \geq 0$.  For $\lambda < \sqrt{n}$
the scalar energy will still red-shift slower than the other
component and it will always come to dominate asymptotically,
approaching the attractor (\ref{eq: jhsoln}). For
$\lambda > \sqrt{n}$, however, the scalar field energy 
cannot always scale as in the case $\rho_n=0$. By doing
so it would become arbitrarily sub-dominant relative to the
component $\rho_n$, and thus arbitrarily strongly damped.
Eventually this damping must reduce its kinetic energy
so that it will then scale slowly (since $\xi \rightarrow 1$)
and begin to catch up again with the $\rho_n$ component. 
It is not surprising then to find that there are in fact, 
for $\rho_n \neq 0$, a quite different set of solutions 
to (\ref{eq: potleoma})-(\ref{eq: potleomc})
for the exponential potential\cite{RP,wett},
in which the energy in the scalar field mimics that of the 
dominant component, contributing a fixed fraction of the energy 
density determined by $\lambda$ given by
\begin{equation}
\Omega_\phi\equiv \frac{\rho_\phi}{\rho_\phi + \rho_n}= \frac{n}{\lambda^2}
\qquad \rho_\phi \propto 1/a^n  \qquad
\xi =1-\frac{n}{6}
\label{attractor}
\end{equation}
This solution is also an attractor \cite{clw}, with $\phi$ again evolving
logarithmically in time as given by (\ref{eq: jhsoln}) (with
only $\phi_o$ differing). 
Fig \ref{transient} 
illustrates the evolution towards the attractor starting from 
initial conditions with the scalar field energy very dominant (obtained 
by a numerical evolution of the homogeneous equations of motion for 
the exponential scalar field with $\lambda=4$ and components of
radiation and matter which are equal at $a=1$).
The scalar field energy first scales rapidly, approximately as 
$1/a^6$ as indicated by the solution (\ref{eq: jhsoln}), until 
it has undershot the energy density in the radiation. 
It then turns around and starts scaling much slower than radiation
or matter until it again catches up with them in the matter
dominated regime, and then settles down at the fraction given by
(\ref{attractor}) with $n=3$.
As anticipated above the main feature - the turn around in 
the scaling - can be understood to arise from the increase in damping
as the radiation/matter becomes dominant. One can easily see quantitatively 
how this comes about: The energy scaling as $1/a^6$ scaling
occurs while the $V'(\phi)$ term in (\ref{eq: potleoma})
is very subdominant and we have then $\dot{\phi} \propto 1/a^3$.
If the dominant component on the RHS of (\ref{eq: potleoma}) is 
radiation/matter one then obtains the evolution of the field
\begin{eqnarray}
\phi(t)= \phi_o +\dot\phi_o t_o
\big( 1- (\frac {t_o}{t})^{2}\big)
\qquad \rho_n << \rho_{\phi} \quad (n=3) 
\nonumber\\
\phi(t)= \phi_o +2 \dot\phi_o t_o
\big( 1- (\frac {t_o}{t})^{\frac{1}{2}}\big)
\qquad \rho_n << \rho_{\phi} \quad (n=4)
\label{eq: phiradn}
\end{eqnarray}
In contrast to the case of the logarithmic dependence on time
of the attractors, the potential term evolves in either case
more slowly than the first two terms in (\ref{eq: potleoma}).
This results in a slow-down in the scaling of the scalar energy 
which drives it back eventually towards the dominant component.
We will discuss in greater detail below what determines the duration 
of this transient period in which the scalar energy is very sub-dominant.
We will also see below that the re-entry in this example in the 
matter-dominated epoch is just a result of the inital conditions.  
For $\lambda >2$ the attractor exists both in the 
radiation and matter dominated epochs. Given that the relative  
contribution of the scalar field energy and the dominant component
differs so little in the two epochs (by a factor of $3/4$) 
one would anticipate that the scalar field, if established in the 
attractor in the radiation dominated epoch will match tightly onto 
its asymptotic value in the matter dominated epoch. That this is 
\begin{figure}
\centerline{\psfig{file=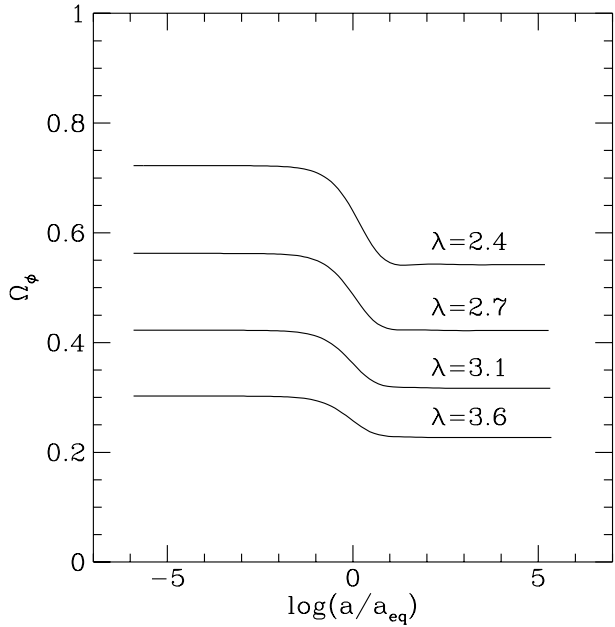,width=3.5in}}
\caption{The evolution of the fractional energy density in the scalar
field for a selection of $\lambda$s}\label{scalfam}
\end{figure}
indeed the case can be seen from Fig \ref{scalfam}, which shows the 
evolution of the fractional contribution to the energy density for
a range of values of $\lambda$ (with the scalar field prepared
in the attractor at the beginning of the evolution).

The conclusion from this discussion is that, in the
presence of matter or radiation, a scalar potential
which is flatter than a simple exponential with 
$\lambda=\sqrt{3}$ will support modes with energy scaling
slower than matter, and always asymptotically dominate.
Simple exponentials with $\lambda < 2$ have
attractors in the radiation and matter dominated epoch
given by (\ref{attractor}), and those with
$\sqrt{3}<\lambda < 2$ in the matter dominated epoch only.
A scalar field rolling down a potential steeper than these
simple exponential  e.g. $\sim e^{-\mu \phi^2/M_P^2}$
will clearly always decay away asymptotically relative
to radiation/matter. We also saw that oscillating modes 
scale in a way which is independent of what other
component is present, and there are no attractor solutions. 

The attractor solutions for the exponential scalar field 
are therefore a special case. Before embarking on the
full treatment of the cosmology in which the attractor 
(\ref{attractor}) describes the unperturbed limit, we
discuss further why this attractor solution is particularly
interesting from a cosmological point of view: By virtue
of its being an attractor, it avoids the tuning problems
which are inevitably a feature of all other scalar field
cosmologies which have been proposed. First, however, we discuss
briefly another attractive aspect of the required
potentials.

\subsection{Theoretical Origin of Exponential Potentials}\label{exptls}
 
Scalar fields with simple exponential potentials 
occur in fact quite generically in certain kinds of particle physics
theories. Because of the existence of the power-law inflationary
solutions (\ref{eq: jhsoln}) such theories have been
studied in some detail by various authors, with attention focussed
on models and parameters which lead to inflation. We will not
carry out any detailed analysis here of particular theories
which would produce the exact parameters required for the present
case, but limit ourselves to a brief review of such theories
and the observation that the parameter values which we require
are quite reasonable in the context of such theories. 
      
The first are Kaluza-Klein theories in which the fundamental theory
has extra dimensions, which are compactified to produce the four
dimensional world we observe. In the effective four dimensional
theory scalar fields arise which correspond to the dynamical degree
of freedom $L$ associated with the `stretching' of the internal dimensions.
The potential is generically exponential because the kinetic term
has the form $(\dot{L}/{L})^2$, while the potential `stretching energy'
is polynomial in $L$. A simple example \cite{wettnpb,halliwellnpb}
is a six-dimensional Einstein-Maxwell theory where the two extra 
dimensions are compactified to $S^2$ with radius $L$.
Defining a field $\Phi$
\begin{eqnarray}
\Phi= 2 M_P \ln L \label{sfKKdef}
\end{eqnarray}
results simply in four-dimensional Einstein gravity minimally
coupled to this field which then has the potential
\begin{eqnarray}
V(\Phi)\propto
\exp(-\frac{\Phi}{M_P})
[1-\exp(-\frac{\Phi}{M_P})]^2
\end{eqnarray}
In the regime of large $\Phi$ we have 
$V(\Phi)\sim \exp(-\frac{\Phi}{ M_P})$ i.e. effectively 
a simple exponential potential with $\lambda=2$.  Another example studied
in detail in \cite{wettshafiplb,wettnpb,wettshafinpb} is the case 
of a pure gravitational theory with higher derivative terms in an
arbitrary number of dimensions, $D+4$, where spontaneous compactification 
arises because of the higher derivative terms. The effective four 
dimensional theory is in this case considerably more complicated, but 
again includes a scalar field $\Phi$ defined as the logarithm
of the ``radius'' of the compactified D-dimensions which gives, in certain
regimes, a potential driving the dynamics of the zero mode of the 
field of the form of (\ref{eq: potleoma}) with 
$V(\Phi)\sim \exp(-{D} \frac{\Phi}{\sqrt{8 \pi} M_P})$.

Another set of models in which such potentials appear is in supergravity 
and superstring theories \cite{supergravity}. One of the most studied 
is the Salam-Sezgin model with, $N=2$ supergravity coupled to matter 
in six dimensions. It predicts \cite{halliwellattractors}
the existence of two scalar fields $\Phi$ and $\Upsilon$ with a 
potential of the form 
\begin{eqnarray}
V(\Phi,\Upsilon)={\bar V}(\Upsilon)\exp(-\sqrt{2}\frac{\Phi}{{ M}_P})
\end{eqnarray}
For $\Upsilon\gg0$ it corresponds to a potential of the required
form we want, with $\lambda=\sqrt{2}$. Further examples relevant to
inflation are given by the authors of \cite{yokoyama-maeda}, who
find two scalar fields with $\lambda=\sqrt{2}$ and $\lambda=\sqrt{6}$
in and $N=2$ model, as well as in an $N=1$ ten-dimensional
model with gaugino condensation \cite{fail}.

A further class of theories in which such fields arise is in
higher order gravity. There is a conformal equivalence
(see \cite{barrowcotsakis}, and references therein) between
pure gravity described by a lagrangian which is an analytic function
of the scalar curvature $R$ and general relativity plus a scalar field
with a determined potential, and in the presence of matter the metric 
associated with the latter description is the physical one \cite{cotsakis}.
For example \cite{barrowcotsakis} in $d$ dimensions the potential 
in the case of a lagrangian quadratic in $R$ is 
\begin{eqnarray}
V(\Phi)\propto \exp (\frac{d-4}{2} \frac{\Phi}{M_P})
[1-\exp (-\frac{d-2}{2}\frac{\Phi}{M_P})]^2
\end{eqnarray}
and, for $d=4$ and a polynomial lagrangian of the form 
$\sum_{n=1}^k a_n R^n$, it is
\begin{eqnarray}
V(\Phi)=A_1 \exp (-2\frac{\Phi}{M_P})
+ \sum_{n=2}^k A_n \exp (-2\frac{\Phi}{M_P})
[1-\exp (\frac{\Phi}{M_P})]^n
\end{eqnarray}
where the $A_n$ are rational coefficients determined by the $a_n$.
As $\Phi \rightarrow +\infty$ we have $V \sim \exp[ (k-2)\Phi /M_P]$.

All previous analysis of the  cosmological consequences of the 
existence of such scalar fields has focussed on inflation, which
is realized in the case of the simple exponential for 
$\lambda  <  \sqrt{2}$. 
The fact that many of these models can at best give 
$\lambda \lta \sqrt{2}$ rather than the considerably flatter
potentials needed for inflation (to naturally give a large number 
of e-foldings and a nearly flat spectrum of perturbations)  
meant that these theories provided a general motivation 
rather than a realistic model. In \cite{yokoyama-maeda}, for
example, an extra ad hoc damping term was introduced to produce a
more satisfactory model from supergravity motivated models.
In the present context this is not the case: We will see that the 
most interesting range for structure formation is 
$\Omega_\phi \in [0.08,0.12]$ in the matter era
which approximately corresponds to the range $\lambda \in [5,6]$.
Although these values are a little larger than in the simplest models
we have reviewed,  some of the models above clearly can lead to parameters
in this range, and, for example, the number of compactified dimensions 
required to give these values in the first type of theory we reviewed 
is certainly not unreasonable in the context of superstring theory.
In terms of the theoretical origin of the potential we study therefore, 
it is not just the form of the potential which is one which arises
naturally, but it may also be that the precise value of the single
free parameter in this potential is a natural one. Clearly 
further analysis of such models would be required to make a 
stronger statement than this, and to see if the cosmological 
effect of these fields which we study in this paper might ultimately
be used to give us specific hints about fundamental particle physics.

\subsection{Nucleosynthesis Constraints}

In considering the possibility that some significant fraction of the 
energy density of the universe may be in a homogeneous mode of a scalar
field, the earliest constraint comes from nucleosynthesis. Such a 
contribution from a weakly coupled scalar field enters in 
determining the outcome of nucleosynthesis (the primordial densities 
of the light nuclei) only through the expansion rate. Adding a component
increases the expansion rate at a given temperature. The dominant
effect of such a change is in its effect on the ratio of neutrons
to protons when the weak interactions keeping them in equilibrium
freeze-out at a temperature of $\sim 1$MeV. The range of expansion
rates at this temperature compatible with observations is usually
translated into a bound on the number of effective relativistic
degrees of freedom at $\sim 1 MeV$. Taking $\Delta N_{eff}$ to be the
maximum number of such degrees of freedom additional to those of
the standard model (with three massless neutrinos) the equivalent
bound on the contribution from a scalar field is
\begin{equation}
\Omega_\phi(1 MeV) < \frac{7\Delta N_{eff}/4}{10.75+7\Delta N_{eff}/4}, 
\label{nucleo-one}
\end{equation}
(where $10.75$ is number of effective degrees of freedom in the standard
model). 
A wide range of values for the upper bounds on $\Delta N_{eff}$
exists in the literature, the variation being due both to the 
data taken into account and the methods of analysis used.
We will not attempt here to review all the issues involved and
accept one or other bound as definitive (see , for example,
\cite{sarkar-review} for
a full discussion).  The tendency in the last
few years has been towards less restrictive bounds than were generally
thought correct previously. A typical value now used is  a bound of 
$\Delta N_{eff}=0.9$ which is given by various authors \cite{cst},
 or even a more conservative one of  $\Delta N_{eff}=1.5$ by others 
\cite{ks,bs}. This range corresponds here to 
\begin{equation}
\Omega_\phi(1 MeV) < 0.13-0.2
\label{nucleo-two}
\end{equation}
This is the range of values we will take when we discuss nucleosynthesis
in the rest of the paper. The result for more restrictive 
nucleosynthesis analyses can be read off from (\ref{nucleo-one}).
Note also that we have assumed here that the standard model number of
relativistic degrees of freedom at $1$MeV. The strict experimental 
lower bound on this number is in fact $9$, given that the 
upper bound on the mass of the $\tau$ neutrino 
is $18$ MeV \cite{particle-data}. In the case that the $\tau$ neutrino
is non-relativistic (and decays before nucleosynthesis) the bound
(\ref{nucleo-two}) is changed to $\Omega_\phi(1 MeV) < 0.27-0.33$.

\subsection{Scalar Fields at Late Time and Fine-Tuning}

In modifications of standard inflationary flat  
cosmologies \cite{RP,steinhardt,liddleviana} involving a contribution from a scalar field 
which have been considered to date, attention has been focussed 
exclusively on the case that the scalar field contributes 
significantly only at recent epochs, at the earliest well 
after the transition to matter domination.
The main reason for this is that one of the motivations for many
models has been to produce a contribution at late times which
scales slower than matter and dominates asymptotically,
producing effects very similar to that of a cosmological constant.
In this case there is unavoidably the same sort of tuning as involved
in the cosmological constant model: One requires an initial
energy density in the scalar field which is characterised by
the energy density in the universe at recent epochs.

Another kind of model which has been considered 
is that in which the energy density in the scalar field scales like
matter asymptotically, implemented in an oscillating
mode of a sinusoidal potential. If the field lies initially
away from the minimum, it becomes important once the curvature
of the potential is comparable to the expansion rate and
initially behave like a cosmological constant before rolling
down the potential and scaling asymptotically like matter.
This model also involves the same sort of tuning, since 
a small energy scale must be introduced to single out this
late time at which the scalar field becomes dynamically important.
It has been argued in \cite{fhsw} that there are particle physics
motivations for the introduction of such a potential with such 
a characteristic scale, and that there is nothing `unnatural' 
about models of this type.

The special case we have discussed of the attractor in the
exponential potential is quite different, simply because it is
an attractor. In contrast the only attractor in the
case of a field asymptotically scaling slower than matter is the
pure scalar field dominated cosmology, and the tuning referred to 
is just that required to make the coupled system of
scalar field and matter approach this attractor just at the 
present epoch. For the models with an oscillating mode 
there is no attractor, and the tuning  consists in fixing the energy 
density in the scalar field to be comparable to that in matter
when it starts scaling like matter. With the attractor for matter
and radiation in the present case, the scalar field plus matter/radiation 
will always end up in this solution asymptotically. In this paper
we will take this solution to apply from the beginning
of our simulation of structure formation, deep in the
radiation era at a red-shift of $z \sim 10^7$ when the temperature
is $\sim 100 eV$. We now address the question as to what range of initial
conditions on the scalar field in the early universe   
will give rise to this behaviour. 
In the course of this analysis we will also determine what
sorts of initial conditions are compatible with the late
entry (in the matter era at a red-shift of $\sim 70$) to
the attractor discussed in \cite{liddleviana}.

\subsection{Initial Conditions in the Early Universe
and the `Self-Tuning' Scalar Field} \label{ics}

The assessment of what are `natural' initial conditions for
a scalar field in the early universe requires of course a 
particular framework within which to address the question.
What we need to determine is essentially just the kinetic and
potential energy in the scalar field at some early time relative to 
that in radiation (and therefore matter). Given that we are working in the
context of inflation-motivated flat homogeneous cosmologies
(and will assume Gaussian adiabatic perturbations of the type produced
by inflation) we assess the question within this framework.
The energy density in radiation is then determined by how
the universe is reheated after inflation. We consider both the
usual scenario for reheating by decay of the inflaton and
then an alternative scenario introduced in \cite{spokoiny}. The
reason for our detailed analysis of this second non-standard case
will become clear below.

First consider the standard reheating scenario. We suppose
there is an inflaton field and the scalar field with an
exponential potential $V_oe^{-\lambda \phi/M_P}$ with
$\lambda >2 $ (i.e. with the attractor in the radiation/matter
epochs which we will consider). Let us consider a typical
inflationary model e.g. chaotic inflation.
The simplest and most natural assumption for the relative energy 
densities at the onset of inflation is
\begin{equation}
\frac{1}{2} \dot \phi^2 \sim V(\phi)\lta  V_{inf}
\label{bcs-inflation}
\end{equation}
where $V_{inf}$ is the energy density in the inflaton.
We make the latter assumption since it is required for the 
onset of inflation.
The dynamics of the two fields are then described by
the equations (\ref{eq: potleoma})-(\ref{eq: potleomc}), but
with $\rho_n$ now the total energy of the inflaton and $n$
a function of the inflaton $n=6(1-\xi_{inf})$, where 
$\xi_{inf}$ is the ratio of the potential energy of the 
inflaton to its total energy. For the same reasons 
as in the case discussed above when $\rho_n$ describes matter or 
radiation, there is an attractor solution given by
(\ref{attractor}) with $n=6(1-\xi_{inf})$ (assuming $n$ changes 
slowly which is the case for slow-roll inflation). 
The energy in the scalar field scales so slowly because
its roll is strongly damped by the inflaton, with
(\ref{attractor}) specifying the exact ratio of energies
at which the damping slows the roll to give precisely the
same scaling as for the inflaton. 
In inflation driven by an exactly constant energy density, this 
ratio is zero and just represents the asymptotically approached 
solution in which the scalar field rolls away to an arbitrarily large 
values of the field after an arbitrarily long time. 
In any realistic model of inflation however the inflaton 
must roll in a non-trivial potential in order to exit from 
inflation and $\xi_{inf} \neq 1$.  For example, in
chaotic inflation in a potential $\sim \phi_{inf}^4$,
one has, in the slow-roll inflationary regime, 
$n=\frac{8}{3} (\frac{M_P}{\phi_{inf}})^2$.  Once inflation
commences, say at $\phi_{inf}=\phi_{inf}^o$, the energy in the 
scalar field will be driven towards 
\begin{equation}
\Omega_\phi^o = \frac{n_o}{\lambda^2}= \frac{8\Omega_\phi^m}{9} 
\big(\frac{M_P}{\phi_{inf}^o}\big)^2 \approx \frac{\Omega_\phi^m}{9N_e}
\label{inflation-fraction}
\end{equation}
where $\Omega_\phi^m$ is the fraction of the energy density in the
attractor in the matter era, and 
$N_e \approx \frac{1}{8}(\frac{\phi_{inf}^o}{M_P})^2$
is the number of e-foldings of inflation.
As the inflaton rolls down the potential $n$ increases
and the fraction of energy in the attractor grows.

Starting from initial conditions like (\ref{bcs-inflation})
the roll of the scalar field is very rapidly over-damped
due to the rapid red-shifting of its initial energy  density. 
It will always then scale much slower than it would in the
absence of the inflaton, with $\xi \approx  1$. 
In the next section we will see that this is enough to ensure 
that its energy density relative to the inflaton will never drop 
substantially below that in the attractor. (The transient with
sub-dominance we observe in that case results from the field
having initially evolved  without the $\rho_n$ component playing
any role for a long period). Without analysing in detail  
how precisely the energy density can track the (growing) attractor 
value during inflation, we thus conclude that the fraction of the 
energy in the scalar field at the end of inflation will be bounded 
below by $\Omega_\phi^o$ as given by (\ref{inflation-fraction}). 

\begin{figure}
\centerline{\psfig{file=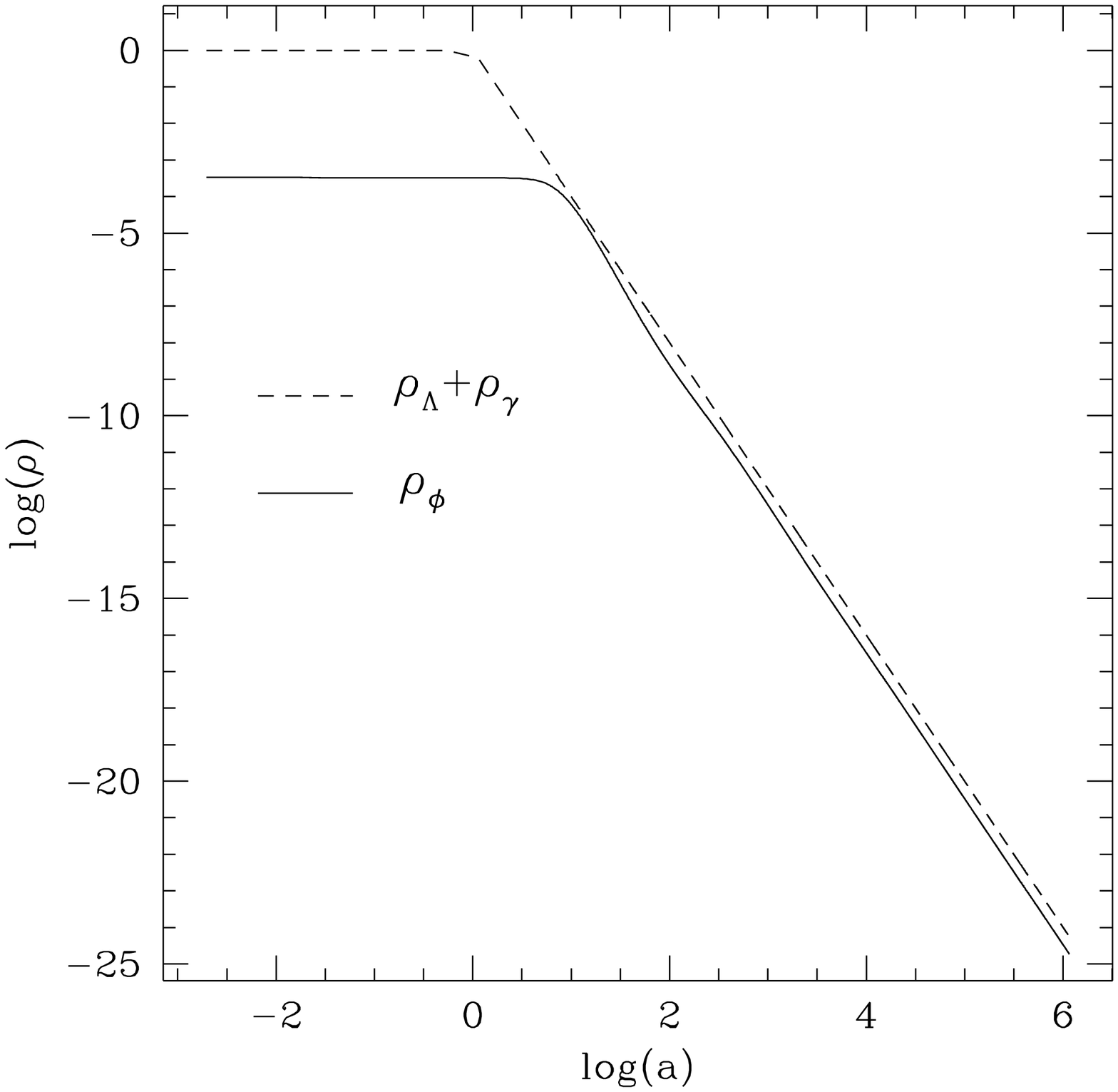,width=3.5in}
\psfig{file=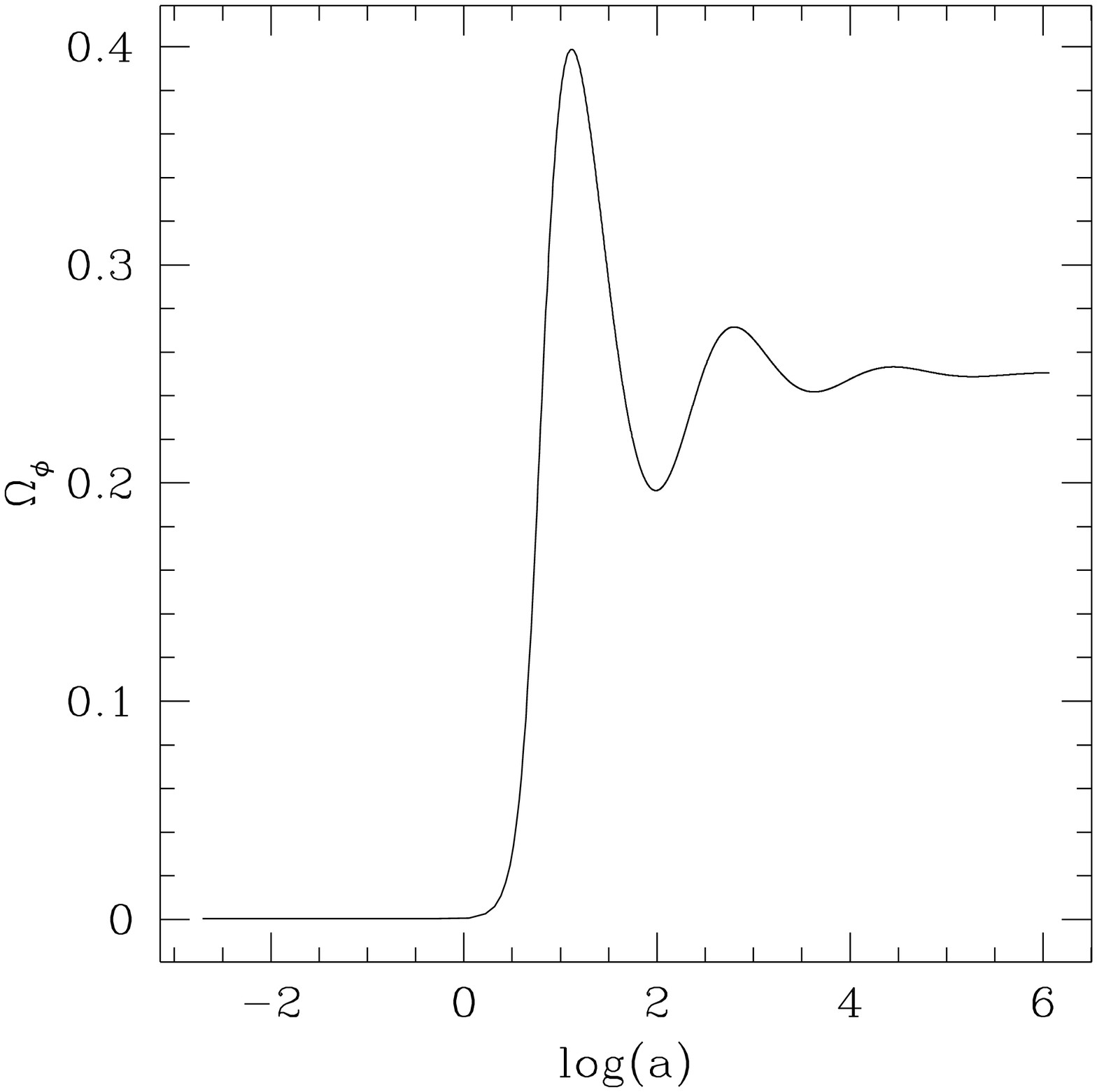,width=3.5in}}
\caption{In the left panel we plot the evolution of the energy density
in the scalar field ($\rho_\phi$) with $\lambda=4$ and in the remaining matter as a
function of scale factor 
(the units are arbitrary), in the case that the scalar field 
is initially sub-dominant.
In the right panel we plot the evolution of the
fractional density in the scalar field.}\label{inflation}
\end{figure}

From Fig \ref{inflation} (and our previous discussion of
Fig \ref{transient}) we see that this conclusion is enough 
to establish that, after inflation, the attractor for
the scalar field in the radiation dominated epoch is approached 
within a few expansion times of the end of inflation.
When the inflaton starts oscillating and/or decays it begins to
scale like matter or radiation, while the scalar field remains
overdamped and its energy almost constant until the 
attractor is attained. In Fig \ref{inflation} we assume an
abrupt transition from inflation with essentially constant
energy to radiation type scaling at scale factor $a=1$,
and take the initial fraction of the energy in the scalar field 
to be $3 \times 10^{-4}$. The energy density approaches that
in the attractor by $a \sim 10$, and then oscillates about it
as it approaches it asymptotically. Without further detailed
analysis it is clear that for a typical inflationary model,
in which the reheat temperature is at most a few orders of
magnitude below the GUT scale, the attractor will be approached
to great accuracy not only by the time scales relevant to structure 
formation, as we assume below, but also long before nucleosynthesis.

Late entry into the attractor for the scalar field, in 
particular late in the matter dominated epoch as assumed 
in the case of this type treated in \cite{liddleviana}, is
therefore certainly not what one would expect from typical
inflationary models with the usual method of reheating.
Inflation is however not embodied in a specific model
or set of models, and it would certainly be possible (and easy
to see how) to devise a model in which the energy density
in the scalar field at the end of inflation could be tuned
to any required value. 
 
Given that the attractor solution is typically established 
by nucleosynthesis, one must satisfy the constraint
discussed above. Converting (\ref{nucleo-two}) to one 
on the contribution from the scalar field  at late times
(in the matter dominated epoch) gives\footnote{The upper
bound here which corresponds to the analysis of \cite{ks}
is actually given explicitly in this form (for the 
exponential attractor) in \cite{bs}.}
\begin{equation}
\Omega_\phi < 0.1 -0.15 \qquad {\rm or} \qquad \lambda > 5.5 - 4.5
\label{nucleo-constraint}
\end{equation}
It is because this quantity would seem to be too small to
play any important cosmological role that this kind of
attractor solution has been disregarded by the authors 
who have noted these solutions\footnote{
Ratra and Peebles \cite{RP}  give this explicitly as the reason for
the attractor solution model being ``phenomenologically untenable'' (p. 3416).
Wetterich in \cite{wett} considers the homogeneous cosmology of 
the case in which the exponential field dominates at late times; in
order to satisfy the nucleosynthesis constraint he discusses the
possibility of late entry brought about by a tuning of the 
initial scalar energy density to an appropriately small value, and  
also the possibility that  $\lambda$, rather than being constant, 
decreases between nucleosynthesis and late times.}.  
That this is  not in fact the case we will see in greater detail 
in the later part of this paper. Because the contribution 
is important for a long time it can have just as significant an effect
as a larger contribution which plays a role only at late times.

The initial motivation for our detailed study of this model
was not, however, the realization that the attractor might 
consistently be established by nucleosynthesis for cosmologically
relevant cases. Rather our starting point was the observation
that there are interesting and viable cosmologies in which 
scalar field energy in a rapidly scaling mode could dominate 
over radiation in the very early universe, and that in the case 
that this field is an exponential of the type relevant to late 
time cosmology, there may be, as shown in Fig \ref{transient} 
a transient period between the two epochs (of domination by the 
scalar field, and the late time attractor) lasting many 
expansion times in which the scalar field energy is negligible.
If this transient period includes nucleosynthesis the constraint
(\ref{nucleo-constraint}) would not apply. We now discuss this model
and examine how the time of entry into the attractor 
depends on the parameters in the model. 

\subsection{Late Entry in an Alternative Model of Reheating}

An epoch dominated by a scalar field in a mode scaling
faster than radiation (or {\it kination} \cite{mj})) comes 
about by construction in an alternative theory of reheating 
suggested in \cite{spokoiny}. Instead of rolling down to the minimum
of a potential, oscillating and decaying, as envisaged in 
the standard reheating scenario, the inflaton field can
roll into a steeper potential supporting a mode scaling
faster than radiation (i.e. an exponential with $\lambda >2$
or steeper). Rather than being, as is often stated, completely 
`cold and empty' at the end of inflation, the universe contains
a component of radiation created simply by the expansion of the
universe (with energy density $\sim H^4$ ). Although initially
very sub-dominant relative to the energy $\rho_\phi$
in the scalar field  ($H^4/\rho_\phi \sim \rho_\phi/M_P^4$)
 a transition to a radiation dominated cosmology will 
take place at some subsequent time since the radiation 
red-shifts away slower than the energy in the scalar field. 

We restrict ourselves to the case that the relevant field
is the simple exponential with $\lambda > 2$ (i.e. 
$\Omega_\phi < 0.75$). We evolve the system of scalar 
field plus radiation and matter forward in time from
the end of inflation, at which time we take the initial
conditions to be specified by 
\begin{equation}
V(\phi_o) = 3M_P^2 H_i^2 \quad \dot{\phi}_o=0 \quad 
\rho_{\rm rad}^o= \epsilon H_i^4 \quad \epsilon=10^{-3}
\label{initial}
\end{equation}
where $H_i$ is the expansion rate at the end of inflation.
The initial condition assumes an abrupt end to slow-roll
inflation i.e. when the inflaton begins rolling in
the region in which the potential is exponential, the
potential energy is still dominant. The choice  
$\epsilon=10^{-3}$ corresponds to the simplest
estimate of the initial radiation density, taking it to be 
dominated by the radiation at the horizon scale at the end
of inflation (with `temperature' $T \sim H/2\pi$ \cite{brandenberger},
see \cite{spokoiny,mjtp} for a more detailed discussion).
We will see below that the results we are interested in here
are not very sensitive to these choices.
 
In this model with the exponential potential there are 
then just two parameters: $H_i$, the expansion rate at the 
end of inflation, and $\lambda$ (or, equivalently, 
$\Omega_\phi\equiv 3/\lambda^2$ in the attractor in 
matter domination). These entirely specify the
post-inflationary homogeneous cosmology, which evolves,
as illustrated in Fig \ref{transient}, from the
scalar field dominated phase through a transient
into the late time attractor. What we want to determine
here is the time at which the attractor is approached
as a function of $H_i$, and the range of parameters
for which the model is compatible with nucleosynthesis.

\begin{figure}
\centerline{\psfig{file=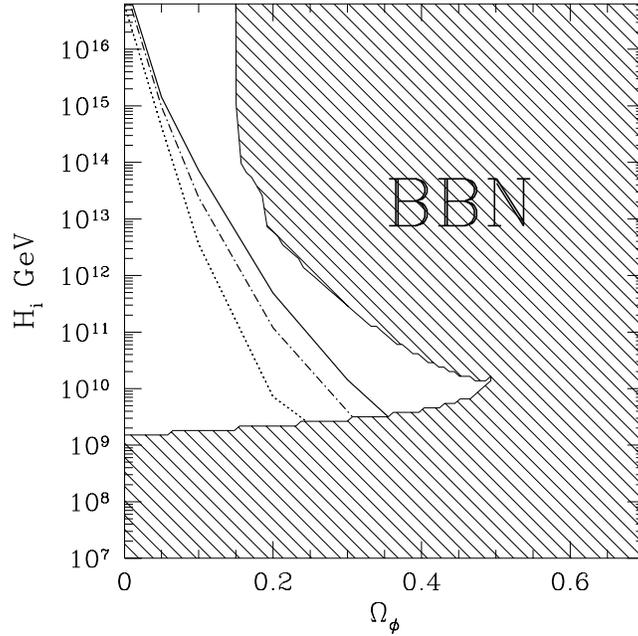,width=3.5in}}
\caption{Reheating by kination in a simple exponential potential: 
The solid region is that excluded by nucleosynthesis
constraints. The solid line (dash-dot, dotted) show the models for 
which the attractor is established at the beginning of structure 
formation (matter domination, today).}\label{HvsOmega}
\end{figure}
  
The results, which are summarized in Fig \ref{HvsOmega}, 
can be understood after some closer examination of the
solutions. The first feature to note is that the nucleosynthesis
constraint, denoted by the hatched region, allows a range 
of $H_i$ at given $\Omega_\phi$ which is (i) only bounded below
for $\Omega_\phi < 0.15$, (ii) bounded above and below for
$\Omega_\phi > 0.15$ in a range which pinches off
as $\Omega_\phi \rightarrow 0.5$, so that for larger values
of $\Omega_\phi$ there is no parameter space in the model
compatible with nucleosynthesis. 

This can be understood as follows. The lower bound 
comes from the requirement that the initial regime
of scalar field dominance end sufficiently long before
nucleosynthesis. In its initial phase the scalar field 
approaches rapidly a mode in which it scales approximately as
$1/a^6$ for $\Omega_\phi < 0.5$ (or as given by 
(\ref{eq: jhsoln}) for $\Omega_\phi < 0.5$ ($\lambda > \sqrt{6}$)). 
In order that the radiation $\rho_{rad}^i$ come to dominate 
by the time its temperature ($T \sim H_i a_i/a$) corresponds
to that at nucleosynthesis ($T_{ns}\sim 1$ MeV), this means 
(from (\ref{initial})) that
\begin{equation}
H_i > M_P \big( \frac{T_{end}}{M_P} \big)^{\frac{1}{2}}
\big( \frac{3}{\epsilon} \big)^{\frac{1}{4}} \approx 10^7 {\rm GeV} 
\label{lower-bound}
\end{equation}
where $T_{end}$ is defined as the temperature at which the
scalar field energy becomes equal to that in the radiation,
and we took $T_{end} = 3$ MeV to obtain the numerical value. 
The higher value of the lower bound in Fig \ref{HvsOmega}
results from the fact, since our initial
conditions are still inflationary,
the scalar field takes a short time to attain the 
rapidly scaling behaviour assumed in deriving
(\ref{lower-bound}). The increase of the lower bound with 
$\Omega_\phi$ is explained in the same way - the transition 
to the $1/a^6$ scaling takes longer as the potential flattens 
(as $\lambda$ decreases). From (\ref{lower-bound}) we also
can see quantitatively the weak dependence on $\epsilon$. 

\begin{figure}
\centerline{\psfig{file=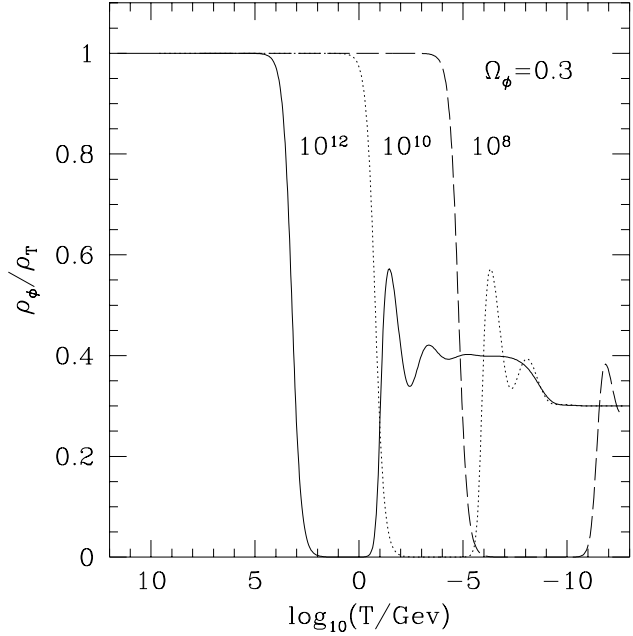,width=3.5in}}
\caption{The fractional energy density as a function of temperature for
$\Omega_\phi=0.3$ and three values of $H_i$.}\label{transfam}
\end{figure}

\begin{figure}
\centerline{\psfig{file=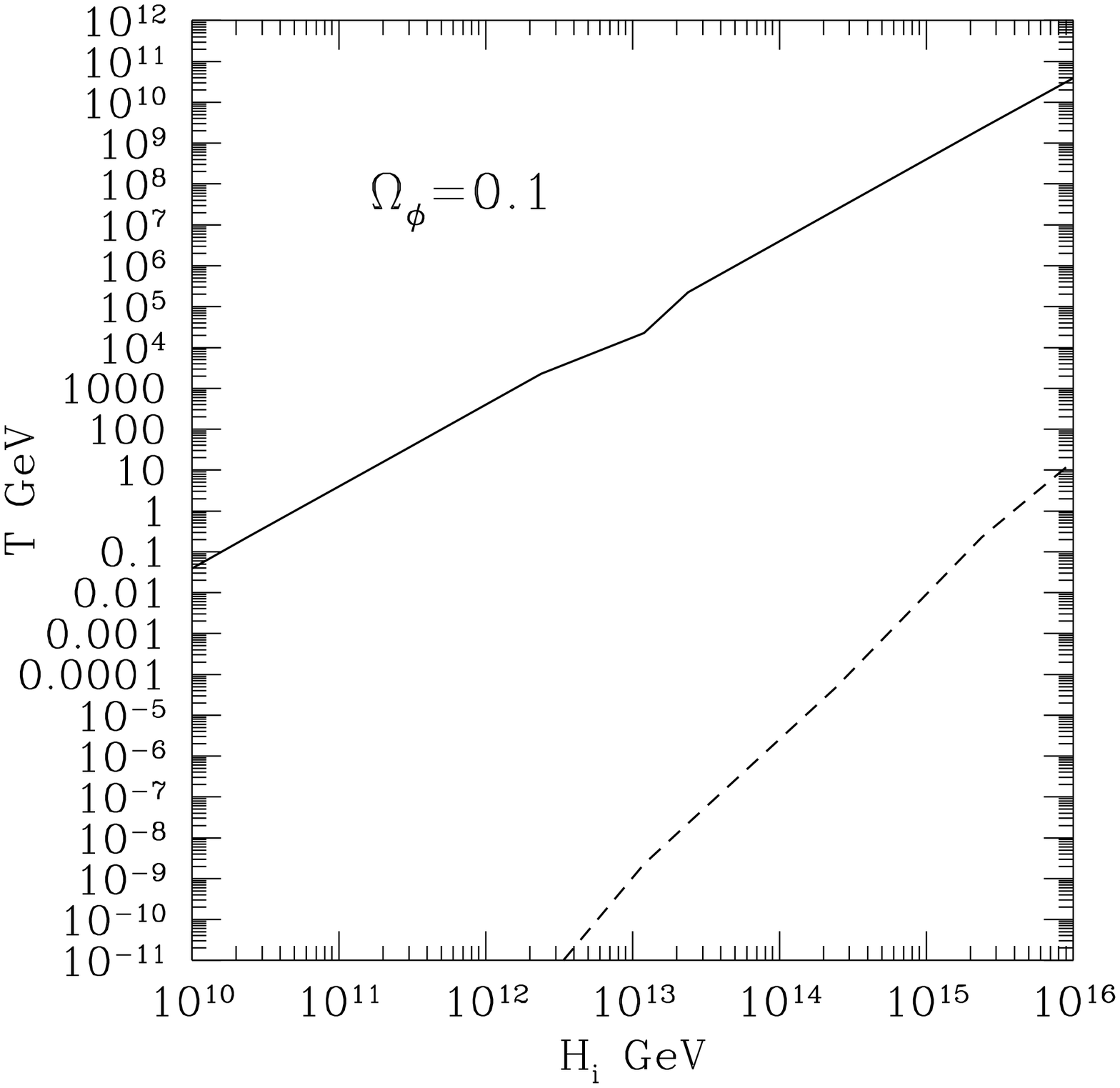,width=3.5in}
\psfig{file=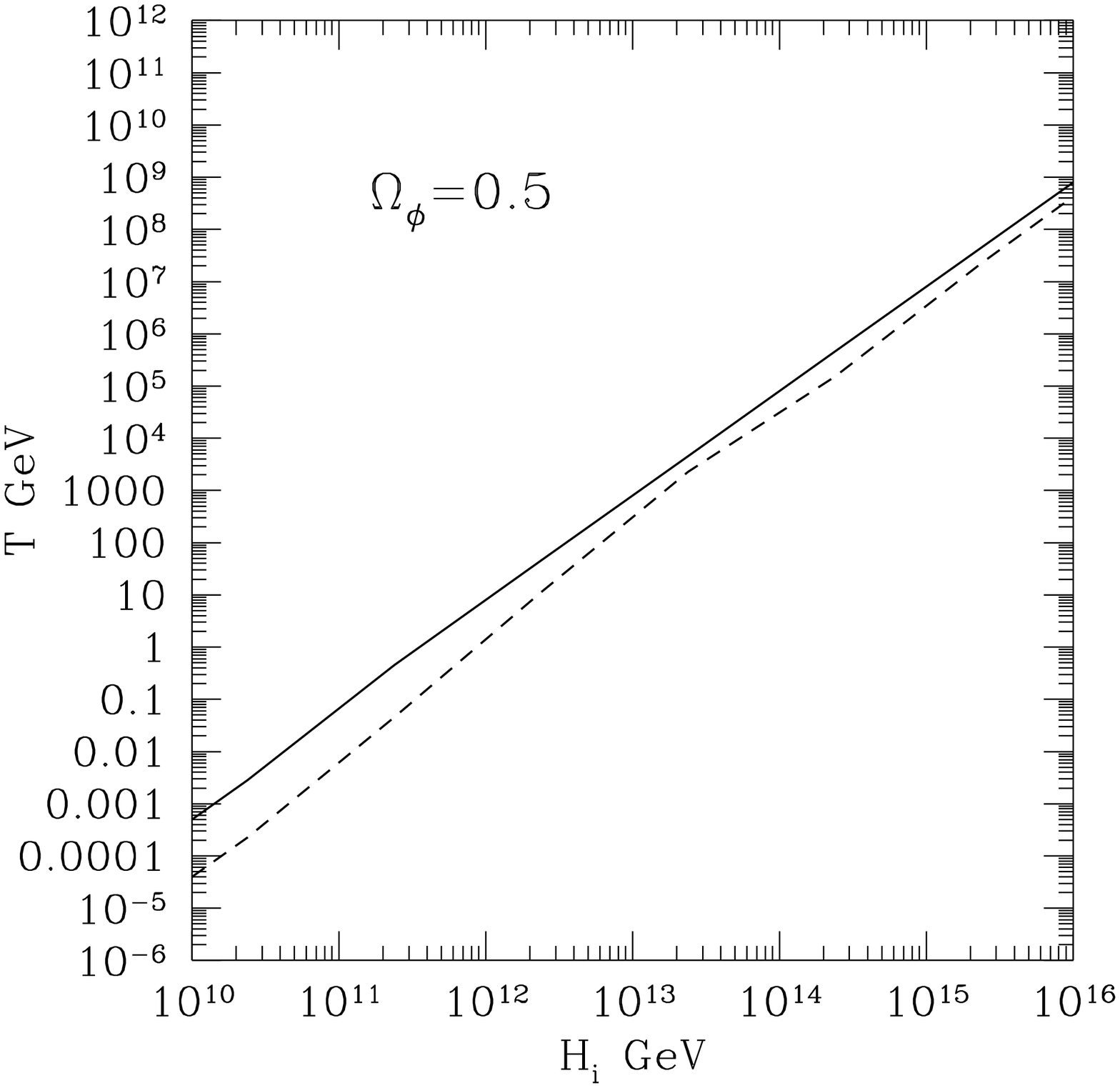,width=3.5in}}
\caption{The solid line is the temperature at
which the scalar field enery density first equals that in
the radiation and the dashed line is the temperature at which
the attractor is established.}\label{temp}
\end{figure}

As we observed in Fig \ref{transient} the scalar field dominance
is followed by a transient period in which it is very sub-dominant,
before approaching the late-time attractor. The upper bound 
on $H_i$ which appears at $\Omega_\phi=0.15$ is just the requirement
that, for $\Omega_\phi > 0.15$ (using the weaker limit of the bound 
in (\ref{nucleo-two})), this period lasts until after nucleosynthesis. 
This gives an upper bound because the transient ends earlier according
as the first phase of scalar field dominance does. In Fig \ref{transfam}
we see the evolution of the fraction of the energy density
as a function of temperature for various different $H_i$. 
For $H_i=10^{12}$ GeV the nucleosynthesis bound is violated because
we enter the attractor too early ; for $H_i=10^{8}GeV $  it is 
violated because we exit the first phase too late. Why the allowed range of
$H_i$ `pinches off' as we go to $\Omega_\phi=0.5$ is that the
duration of the transient decreases as we approach this value.
This can be seen clearly from Fig \ref{temp} 
 which show, for $\Omega=0.1$ (left panel)
and $\Omega=0.5$ (right panel) the temperature $T_{end}$ (when the
scalar field energy first equals that in the radiation)
and at $T_{att}$, when the attractor is established.
This behaviour can be understood by looking back to the
attractor solutions (\ref{eq: jhsoln}). In the late time 
attractor the parameter $\xi =V(\phi)/\rho_\phi$ must reach 
a certain determined value; as indicated by (\ref{eq: jhsoln})
the value $\lambda=\sqrt{6}$ separates two quite distinct regions
in which $\xi$ behaves quite differently. For $\lambda <\sqrt{6}$
the $\rho_n=0$ attractor tends to a finite value; in the
steeper potential with $\lambda > \sqrt{6}$ the potential energy
`runs away' faster and $\xi$ decreases rapidly towards zero.
As long as $\xi \approx 0$ the scalar energy continues
to scale away as $1/a^6$, and it may take a long time for the
kinetic energy to catch up again on the potential energy
(which essentially stops decreasing once the field's roll
is damped by the radiation or matter). The large `undershoot'
and long period of sub-dominance of the scalar energy thus results
from the fact that the exponential field with $\lambda > \sqrt{6}$
evolves for a long time damped only by its own energy density,
which allows the field to run away so fast that the
potential energy decreases enormously.
On the other hand for $\Omega=0.6$ ($\lambda=\sqrt{5}$)
the scalar field in the first phase follows the 
$\rho_n=0$ attractor (\ref{eq: jhsoln}) with 
$\xi=5/6$, and only a little adjustment is needed to find 
the late time radiation dominated attractor with $\xi=2/3$.
The nucleosynthesis constraint is therefore never satisfied
in this case.

Also shown in Fig \ref{HvsOmega} are the curves corresponding to
`entry' to the attractor at different temperatures - at $100$eV
(solid line), at matter-radiation equality (dash-dot line)
and today (dotted line). The undashed region lying to the right
of the first of these lines corresponds to the parameter space 
(for this model) we will consider in the rest of this paper,
which is consistent with nucleosynthesis and with our assumption
of the validity of the attractor with (\ref{attractor}) as
our homogeneous solution.  The line corresponding to re-entry at
nucleosynthesis is not given, but one can infer that it would 
mark out a substantial part of this region (bounded by the
solid line), which corresponds to entry into the the attractor 
after nucleosynthesis but prior to $T=100eV$. There is also a 
region, bounded by the first and third line, which we do not 
describe and which might be of relevance to late time cosmology. 
The region to the left of the dotted line corresponds to regions 
where the scalar field would still be irrelevant today, and 
therefore the existence of the exponential field has no 
implications for late time cosmology.

One can also determine from Fig \ref{HvsOmega} that, for entry
in the matter dominated epoch, the largest possible value of
$\Omega_\phi$ in this model is approximately $0.3$. This places 
an upper bound on what contribution to the energy density might 
be expected to come from this type of solution, if one is not to use the 
sort of tuning which, as we have emphasised, the model has 
the merit of being able to avoid. (In particular we note
that we cannot obtain the case $\Omega_\phi =0.6$ of \cite{liddleviana}.)

Is there a `natural' value for $H_i$ within the allowed range?
$H_i$ is related to the inflaton energy at the end of inflation
by $\rho_i \sim (H_i M_P)^{2}$, and the range 
$H_i \in [10^{8},10^{16}]$GeV corresponds approximately
to $\rho_i^{1/4}\in [10^{13},10^{17}]$GeV.  
With a more developed model with a full ansatz for
the part of the potential which supports inflation, one
can relate $H_i$ to the amplitude of density perturbations,
and require that they match those observed in the CMBR by COBE.
The simplest calculation (for a very flat inflaton potential)
would give $\rho_i^{1/4} \sim 10^{16}$ GeV,
or $H_i \sim 10^{14}$GeV. However the result is not
model independent. In \cite{mjtp} a particular model
is constructed in which $H_i$ can be as low as the 
value at which the phase of scalar dominance ends just before
nucleosynthesis. This parameter range for the model has
the particular interest that it leads to quite radical modifications
of physics at the electroweak scale, since the expansion rate
can be up to five order of magnitude greater than in the
radiation dominated case at $T \sim 100 GeV$.

In the rest of this paper we study only cosmology starting
from $T \sim 100 eV$. We will assume the existence of the
simple exponential, and take the homogenous cosmology about
which we perturb to be given exactly by the attractor
at this initial time. $\lambda$ is thus the sole adjustable
parameter additional to those of SCDM. To allow the simplest
direct comparison of structure formation in our model
with SCDM we take the fiducial value $\Omega_bh^2=0.0125$ 
for the baryon fraction. In the present study we will
not consider the effect of varying this parameter. If
we were to do so we would have to be more definite about our
assumptions about nucleosynthesis, since the allowed range of
$\Omega_bh^2$ (or equivalently baryon to photon ratio $\eta$)
depends on the expansion rate at nucleosynthesis.
The canonically quoted range, which  corresponds to that 
consistent with observations for the case $N_\nu=3$, narrows as 
the expansion rate at freeze-out increases (since the fraction of 
Helium produced increases as the expansion rate does).
As the expansion rate reaches its upper bound,  the $\eta$ required 
for consistency with observations is pushed towards its lower bound.   
Having used the constraints of structure formation to determine the best value
of $\Omega_\phi$ in our model, we will comment again in our conclusions
on the consistency of entry to the attractor prior to nucleosynthesis. 

\section{Evolution of Perturbations}\label{pert}

We now analyse the evolution of perturbations in this
cosmology. The structure of this section is as follows. 
We firstly write down the equations for the unperturbed 
cosmology in conformal coordinates which we use in the rest
of the paper. For completeness,
in \ref{eqpert} we present the equations for the evolution 
of cosmological perturbations in the synchronous gauge. In
\ref{pertdens} we derive analytic solutions to the evolution of
perturbations on sub- and superhorizon scales in the radiation and
matter era. It becomes apparent that they are very similar
to those in a cosmology in which part of the matter content is 
in a massive neutrino (MDM). It is thus instructive to 
identify the key differences between our scenario and the
MDM model, and we do so in \ref{pertmdm}.
In \ref{Cell} we look at the evolution of perturbations in
the radiation and how it affects the temperature anisotropy angular
power spectrum. 

\subsection{Equations of Motion and Initial Conditions for the Background}

Normalising the scalar field in units of $M_P$ and defining
the origin of $\Phi$ so that the scalar potential is
$M_P^4 e^{-\lambda \Phi}$, the equation of motion
for the unperturbed background scalar field is,
in conformal coordinates for which $ds^2=a^2[-d\tau^2+\delta_{ij}dx^idx^j]$, 
\begin{eqnarray}
{\ddot \phi}+2{\cal H}{\dot \phi}-a^2 \lambda M^2_{P}
\exp(-\lambda\phi)=0\nonumber \\
{\cal H}^2 = \frac{1}{3M_{P}^2}a^2(\rho_\phi + \rho_M) 
\label{eq: potleomb1}
\end{eqnarray}
where $a$ is the scale factor, ${\cal H}$ is the conformal Hubble
factor, dots denote derivatives with respect to conformal 
time $\tau$, $\rho_M$ is the energy density in matter and radiation,  and 
\begin{equation}
\rho_{\phi}=M^2_{P}\left [{1 \over 2 a^2}{\dot \phi}^2+M^2_{P}
\exp(-\lambda \phi)\right ]
\end{equation}
In the attractor (\ref{attractor})  we have
\begin{equation}
{\rho_K \over \rho_\phi}={n \over 6} \qquad
{\rho_{\phi}\over \rho_{total}}= \frac{n}{\lambda^2} 
\qquad \rho_{total}= 3M^2_{P}\left(\frac{{\cal H}}{a}\right)^2 
\label{attractor-ratios}
\end{equation}
where $\rho_K= {1 \over 2 a^2}{\dot \phi}^2$ is the kinetic energy
density of $\phi$.
We use these to fix the initial conditions on $\phi$ and ${\dot \phi}$
to be  
\begin{equation}
{\dot \phi}=\frac{n}{\lambda}{\cal H} \qquad
\phi=-\frac{1}{\lambda}\log\left(\frac{4n(6-n){\cal H}^2}
{a^2 \lambda^2}\right)\label{scaling2}
\end{equation}
where $n=4$, since we begin the evolution deep in the radiation 
dominated epoch.

\subsection{Linear Perturbation Theory\label{eqpert}}
We use the notation and results of
Ma $\&$ Bertschinger \cite{MB}, with the modifications brought about
by the addition of a scalar field; we present the full set of equations 
but we refer the reader to \cite{MB} and \cite{SZ}
for details on how to solve them. The formalism we work in is 
the synchronous gauge where
$ds^2=a^2[-d\tau^2+(\delta_{ij}+h_{ij})dx^idx^j]$. Restricting
ourselves to scalar perturbations, 
the metric perturbations can be parametrised as
\begin{eqnarray}
h_{ij}=\int d^3ke^{i{\bf k}\cdot{\bf x}}[{{\bf \hat k}_i}{{\bf \hat
k}_j}h({\bf k},\tau)+({{\bf \hat k}_i}{{\bf \hat k}_j}-\frac{1}{3}
\delta_{ij})6\eta({\bf k},\tau)].
\end{eqnarray} 
Pressureless matter (cold dark matter) has only one non-zero
component of the perturbed energy-momentum
tensor:
\begin{eqnarray}
\delta T^0_0&=&-\rho_c\delta_c 
\end{eqnarray}
and it evolves as
\begin{eqnarray}
{\dot \delta_c}=-\frac{1}{2}{\dot h}
\end{eqnarray}
Radiation can be characterised in terms of a temperature brightness
function, $\Delta_T({\bf k},{\bf {\hat n}},\tau)$ and
polarization brightness $\Delta_P({\bf k},{\bf {\hat n}},\tau)$.
These brightness functions can be
expanded in Legendre polynomials of ${\bf {\hat k}}\cdot{\bf {\hat
n}}$:
\begin{eqnarray}
\Delta_T({\bf k},{\bf {\hat
n}},\tau)&=&\sum_{\ell=o}^{\infty}(-i)^\ell(2\ell+1)
\Delta_{T\ell}( k,\tau)P_\ell({\bf {\hat k}}\cdot{\bf {\hat
n}}) \nonumber \\
\Delta_P({\bf k},{\bf {\hat
n}},\tau)&=&\sum_{\ell=o}^{\infty}(-i)^\ell(2\ell+1)
\Delta_{P\ell}( k,\tau)P_\ell({\bf {\hat k}}\cdot{\bf {\hat
n}}). 
\end{eqnarray}
Defining the density, velocity and shear perturbations by 
\begin{eqnarray}
\delta_\gamma=\Delta_{T 0} \ \ \ \ \theta_\gamma=\frac{3}{4}k\Delta_{T
1} \ \ \ \ \sigma_\gamma=\frac{1}{2}\Delta_{T 2}
\end{eqnarray}
the perturbed energy-momentum tensor for radiation is 
\begin{eqnarray}
\delta T^0_0&=&-\rho_\gamma\delta_\gamma \nonumber \\
ik^i\delta T^0_i&=& \frac{4}{3}\rho_\gamma\theta_\gamma\nonumber \\
\delta T^i_j&=&\frac{1}{3}\rho_\gamma\delta_\gamma+\Sigma^i_j \nonumber \\
({{\bf \hat k}_i}{{\bf \hat k}_j}-\frac{1}{3}
\delta_{ij})\Sigma^i_j&=& -\frac{4}{3}\rho_\gamma\sigma_\gamma 
\end{eqnarray}
Thomson scattering couples the radiation and
baryons, and the latter have a perturbed energy momentum tensor:
\begin{eqnarray}
\delta T^0_0&=&-\rho_b\delta_b \nonumber \\
ik^i\delta T^0_i&=& \frac{4}{3}\rho_\gamma\theta_b
\end{eqnarray}
The evolution equations for radiation are:
\begin{eqnarray}
{\dot \delta}_\gamma&=&-\frac{4}{3}\theta_\gamma-\frac{2}{3}{\dot h}
\nonumber \\
{\dot \theta}_\gamma&=&k^2\left(\frac{1}{4}\delta_\gamma-\sigma_\gamma\right)
an_e\sigma_T(\theta_b-\theta_\gamma)\nonumber \\
2{\dot \sigma}_\gamma&=&\frac{8}{15}\theta_\gamma-\frac{3}{5}k\Delta_{\gamma3}
+\frac{4}{15}{\dot h}+\frac{8}{5}{\dot \eta} \nonumber \\& &-
\frac{9}{5}an_e\sigma_T\sigma_\gamma+\frac{1}{10}an_e\sigma_T
(\Delta_{P0}-\Delta_{P2}) \nonumber \\
{\dot \Delta}_{T\ell}&=&\frac{k}{2\ell+1}[\ell \Delta_{T(\ell-1)}
-(\ell+1)\Delta_{T(\ell+1)}]-an_e\sigma_T\Delta_{T\ell} \nonumber \\
{\dot \Delta}_{P\ell}&=&\frac{k}{2\ell+1}[\ell \Delta_{P(\ell-1)}
-(\ell+1)\Delta_{P(\ell+1)}] \nonumber \\ & &+an_e\sigma_T[\Delta_{P\ell}
+\frac{1}{2}(\Delta_{T2} +\Delta_{P0} +\Delta_{P2} )(\delta_{0\ell}
+\frac{\delta_{5\ell}}{5})]
\end{eqnarray}
and for baryons
\begin{eqnarray}
{\dot \delta}_b&=&-\theta_b-\frac{1}{2}{\dot h}\nonumber \\
{\dot \theta}_b&=&-{\cal H}\theta_b+Ran_e\sigma_T(\theta_{\gamma}-
\theta_b)
\end{eqnarray}
$\sigma_T$ is the Thomson scattering cross section, $n_e$ is the
electron density, and $R=\frac{4\rho_\gamma}{3\rho_b}$. 
The evolution equations for massless neutrinos can be obtained
from those for radiation by setting $\Delta_P=R=\sigma_T=0$ and
$T\rightarrow \nu$.

The perturbation $\varphi$ in the scalar field about 
the homogeneous solution $\phi$ has the equation of motion
\begin{equation}
{\ddot \varphi}+2{\cal H}{\dot
\varphi}-{\nabla^2}\varphi+a^2V''\varphi
+{1 \over 2}{\dot \phi}{\dot h}=0 
\end{equation}
and gives rise to perturbations in the energy momentum tensor 
which are 
\begin{eqnarray}
a^2\delta T^0_0 &=&-{\dot \phi}{\dot \varphi}-a^2V'\varphi \nonumber \\
-a^2\partial_i\delta T^0_i &=&{\dot \phi}\nabla^2\varphi \nonumber \\
a^2\delta T^i_i&=&3{\dot \phi}{\dot \varphi}-3a^2V'\varphi.
\end{eqnarray}

Finally we have the perturbed Einstein equations
\begin{eqnarray}
k^2\eta-\frac{1}{2}{\cal H}{\dot h}&=&4\pi Ga^2 \delta T^0_0 \nonumber \\
k^2{\dot \eta}&=&4\pi Ga^2 i{\bf {\hat k}_i}\delta T^0_i \nonumber \\
{\ddot h}+2{\cal H}{\dot h}-2k^2\eta&=&-8\pi Ga^2 \delta T^i_i
\nonumber \\
{\ddot h}+6{\ddot \eta}+2{\cal H}({\dot h}+6{\dot \eta})-2k^2\eta &=&
+24\pi G a^2 ({{\bf \hat k}_i}{{\bf \hat k}_j}-\frac{1}{3}
\delta_{ij})\Sigma^i_j
\end{eqnarray}

Having defined the evolution equations for the perturbations,
all that remains to be specified are the initial conditions.
We are working within the context of inflation
 and consider an initial set of perturbations of the type
generically predicted by the simplest such scenarios: 
We assume  a Gaussian set of  adiabatic perturbations with
a scale invariant power spectrum. This completely
defines the statistical properties of the ensemble.
Putting the adiabatic perturbations in the superhorizon 
growing mode gives \cite{MB}:
\begin{eqnarray}
\delta_\gamma&=&-\frac{2}{3}C(k\tau)^2 \nonumber \\
\delta_c&=&-\frac{1}{2}h=
\delta_b=\frac{3}{4}\delta_\nu=\frac{3}{4}\delta_\gamma
\nonumber \\
\theta_\gamma&=&\theta_b=\frac{14+4R_\nu}{23+4R_\nu}\theta_\nu=
-\frac{1}{18}(k^4\tau^3)\nonumber \\
\sigma_\nu&=&\frac{4C}{3(15+4R_\nu)}(k\tau)^2 \nonumber \\
\eta&=&2C-\frac{5+4R_\nu}{6(15+4R_\nu)}C(k\tau)^2 
\end{eqnarray}
and, as we will see in more detail below, 
\begin{equation}
\delta_c=\frac{5 \lambda }{4}\varphi
\end{equation}
(and $C$ is the overall normalisation).
All the remaining perturbation variables are set to zero initially.

\subsection{Asymptotic solutions to the evolution of density
perturbations \label{pertdens}}
We can arrive at a simple understanding of how perturbations
evolve in this scenario by considering the simplified
case in which ther is only cold dark matter, radiation
and the scalar field. The evolution equations are then:
\begin{eqnarray}
{\ddot \delta}_c+{\cal H}{\dot \delta}_c-\frac{3}{2}
{\cal H}^2(\Omega_c\delta_c+2\Omega_r\delta_r)-2{\dot \phi}{\dot \varphi}
+a^2V'{\varphi}&=&0 \label{simp1} \\
{\ddot \delta}_r+\frac{1}{3}k^2\delta_r-\frac{4}{3}{\ddot \delta}_c&=&0
\label{simp2} \\
{\ddot \varphi}+2{\cal H}{\dot \varphi}+k^2{\varphi}+a^2V''\varphi
-{\dot \phi}{\dot \delta}_c&=&0 \label{simp3}
\end{eqnarray}
where $\Omega_X=\rho_X/\rho_{tot}$ ($X=C,\gamma,\nu,\phi$).

Consider first the superhorizon evolution. In this limit we
set $k^2=0$ and (assuming adiabatic initial conditions as above) 
$\delta_c=\frac{3}{4}\delta_\gamma$. Using the scaling solutions
for the homogeneous mode of the scalar field, in the radiation era
we have:
\begin{eqnarray}
{\ddot \delta_c}+\frac{1}{\tau}{\dot
\delta}_c-\frac{4(1-\frac{4}{\lambda^2})}{\tau^2}
\delta_c-\frac{8}{\lambda \tau}{\dot \varphi}
-\frac{4}{\lambda \tau^2}{ \varphi}&=&0 \nonumber \\
{\ddot \varphi}+\frac{2}{\tau}{\dot \varphi}+\frac{4}{\tau}\varphi
-\frac{4}{\lambda}{\dot \delta}_c &=&=0
\end{eqnarray}
These equations are solvable and give $\delta_c$ and
$\varphi$ as a linear combination of $\tau^2$, $\tau^{-2}$ and
$\tau^{\frac{-1\pm\sqrt{-15+64/\lambda^2}}{2}}$. Taking only the
coefficient of the growing mode to be non-zero, we find (as stated above)
$\delta_c=\frac{5\lambda}{4}\varphi=A\tau^2$.
On  superhorizon scales in the matter era we have
\begin{eqnarray}
{\ddot \delta_c}+\frac{2}{\tau}{\dot
\delta}_c-\frac{6(1-\frac{3}{\lambda^2})}{\tau^2}
\delta_c-\frac{12}{\lambda \tau}{\dot \varphi}
-\frac{18}{\lambda \tau^2}{ \varphi}&=&0 \nonumber \\
{\ddot \varphi}+\frac{4}{\tau}{\dot \varphi}+\frac{18}{\tau}\varphi
-\frac{6}{\lambda \tau}{\dot \delta}_c &=&=0
\end{eqnarray}
which can be solved to give $\delta_c$ and $\varphi$ as a linear 
combination of $\tau^2$, $\tau^{-3}$ and 
$\tau^{\frac{-6\pm\sqrt{-7+12/\lambda^2}}{4}}$.
With the assumption therefore of initial adiabatic perturbations
in the growing mode in the radiation era, we have 
$\delta_c=\frac{28\lambda}{6}\varphi\propto\tau^2$.
We  conclude, therefore, that the evolution of perturbations
on superhorizon scales are exactly the same as in a scalar field
free universe. This is just a manifestation of the fact that, on very
large scales, the dominant interaction is gravitational, which is blind
to matter type. 

Next we turn to the subhorizon evolution of perturbations, i.e.
$k\tau\gg 1$. Consider first the radiation era.
Assuming that the gravitational feedback
is unimportant on small scales, we have
\begin{eqnarray}
{\ddot \delta}_c+\frac{1}{\tau}{\dot \delta}_c 
-\frac{3(1-\frac{3}{\lambda^2})}{\tau^2}\delta_\gamma
-\frac{8}{\lambda \tau}{\dot \varphi}-\frac{4}{\lambda \tau}\varphi
&=&0\nonumber \\
{\ddot \delta}_\gamma+\frac{1}{3}k^2\delta_\gamma&\simeq&0 \nonumber \\
{\ddot \varphi}+\frac{2}{\tau}{\dot \varphi}+k^2\varphi&\simeq&=0
\end{eqnarray}
The last two equations are easy to solve, giving  $\delta_\gamma \propto
e^{\pm\frac{ik}{\sqrt{3}}\tau}$ and $\varphi\propto
\frac{1}{\sqrt{\tau}}J_{\frac{1}{2}}(k\tau)$, 
$\frac{1}{\sqrt{\tau}}N_{\frac{1}{2}}(k\tau)$ where $J_\mu$ and $N_\mu$ are
spherical Bessel functions. Clearly these have little effect
on $\delta_c$ and one can drop these terms from the first
equation to get $\delta_c\propto C_1$, $\log\tau$. The subhorizon
evolution in the radiation era is thus essentially the same in a universe
with and without a scalar field.

In the matter era on sub-horizon scales  the equations reduce
to
\begin{eqnarray}
{\ddot \delta}_c+\frac{2}{\tau}{\dot
\delta}_c-\frac{6(1-\frac{16}{3\lambda^2})}{\tau^2}\delta_c-
\frac{12}{\lambda \tau}{\dot \varphi}-\frac{18}{\lambda \tau^2}\varphi&=&0
\label{simmsub1} \\
{\ddot \varphi}+\frac{4}{\tau}{\dot
\varphi}+k^2\varphi-\frac{6}{\lambda \tau}{\dot \delta}_c&=&0 
\label{simmsub2} 
\end{eqnarray}
To a first approximation we can discard the gravitational
feedback term in Eq. \ref{simmsub2}. The solutions to this 
wave equation are 
$\varphi\propto \frac{1}{\tau^{3/2}}J_{\frac{3}{2}}(k\tau)$, 
$\frac{1}{\tau^{3/2}}N_{\frac{3}{2}}(k\tau)$
i.e. oscillatory solutions with decaying amplitudes. Clearly this will
contribute little to the growing mode and we can drop the last two
terms in Eq. \ref{simmsub1}. The equation is easy to solve and, 
using $\Omega_\phi=\frac{3}{\lambda^2}$ we find
the growing mode solution $\delta_c \propto \tau^{2+\frac{5}{2}(-1+
\sqrt{1-\frac{24}{25}\Omega_{\phi}})}$. The subhorizon growing mode
is therefore suppressed relative to that in a scalar field free universe.

Having derived these approximate forms we now have a rough idea
of what the mass variance per unit in $\ln k$ is, where this is
defined as 
\begin{eqnarray}
\Delta^2(k)=\frac{k^3}{2\pi^2}\langle|\delta_c(k)|^2\rangle.
\label{delta}
\end{eqnarray}
One has (defining $2\epsilon=5(-1+
\sqrt{1-\frac{24}{25}\Omega_{\phi}})$)
\begin{eqnarray}
\Delta^2(k)\propto\left\{\begin{array}{ll} 
k^4 & \mbox{if $k \le  \frac{2\pi}{\tau_{0}}$} \\
k^{4-2\epsilon} & \mbox{if $k \le \frac{2\pi}{\tau_{eq}}$} \\
{\rm constant}+\ln k & \mbox{if $k > \frac{2\pi}{\tau_{eq}}$} 
\end{array}\right .
\end{eqnarray}

\begin{figure}
\centerline{\psfig{file=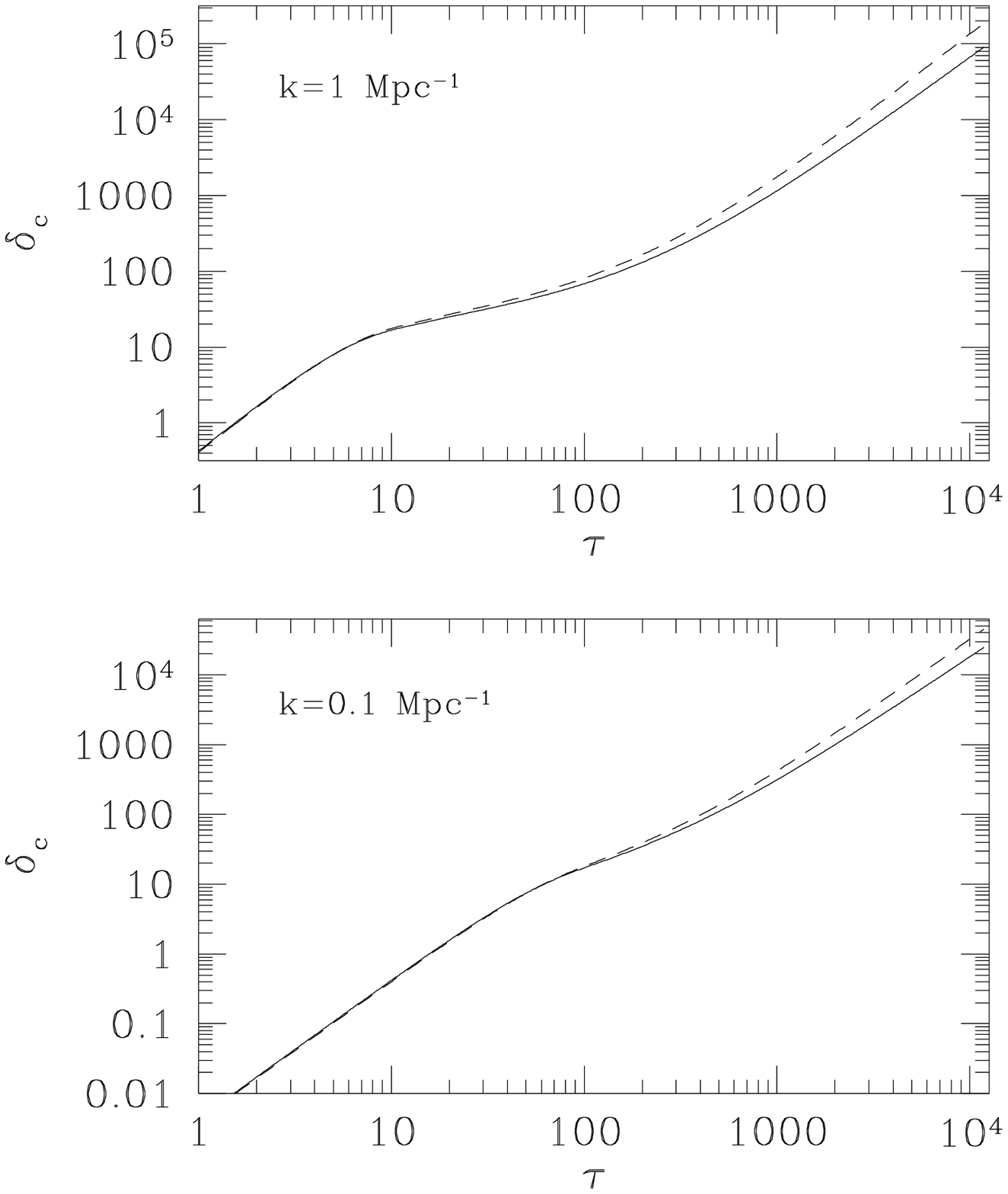,width=3.5in}
\psfig{file=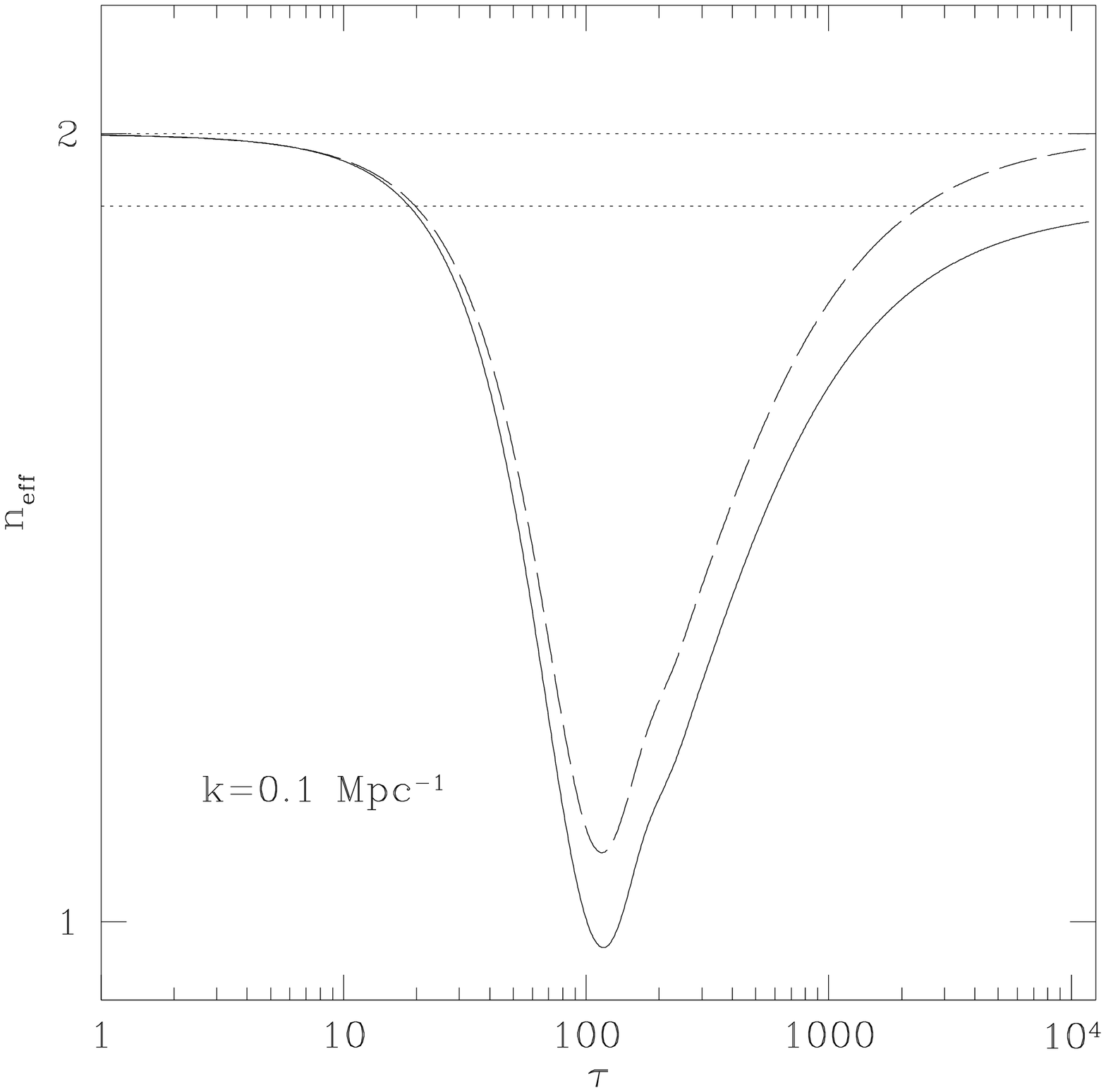,width=3.5in}}
\caption{On the left panel we plot the evolution of $\delta_c$
for two wavenumbers in a sCDM cosmology (dashed line) and in a universe
with $\Omega_\phi=0.1$. In the right panel we plot $n_{eff}$ for
$k=0.1$Mpc$^{-1}$ for the same cosmology. The upper (lower) dotted line
is the asymptotic matter era solution in the sCDM ($\Omega_\phi=0.1$)
case.}\label{deltacphi}
\end{figure}

It is instructive to see the evolution of density perturbations
for a few wavenumbers, one that comes into the horizon
at around radiation-matter equality ($k=0.1$Mpc$^{-1}$) 
and one that comes in during the
radiation era ($k=1.0$Mpc$^{-1}$) for $\Omega_\phi=0.1$. 
They are compared in Fig \ref{deltacphi}
to density perturbations in a universe with no scalar field
and it is clear that there is a suppression of growth after
horizon crossing. Another useful quantity to plot is the dimensionless
growth rate $n_{eff}=\frac{\tau}{\delta_c}{\dot \delta}_c$. For a scalar
field free universe we know that deep in the radiation or matter era
$n_{eff}=2$. For the scalar field cosmology we have deep in the
radiation era $n_{eff}=2$ but deep in the matter era
$n_{eff}=2-\epsilon$. In Fig \ref{deltacphi} we plot $n_{eff}$ for
$k=0.1$Mpc$^{-1}$ and 
can clearly see the different asymptotic regimes. Note that
there is a  long transient to the matter era solution in both
the cosmologies considered.

\subsection{A comparison with the evolution of perturbations with
mixed dark matter\label{pertmdm}}

\begin{figure}
\centerline{\psfig{file=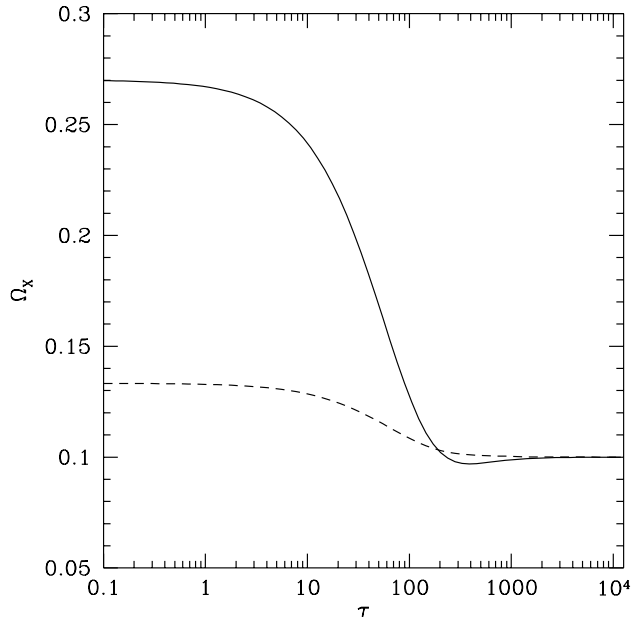,width=3.5in}}
\caption{
The evolution of the fractional energy density in $\phi$ (dashed line)
for a $h=0.5$  universe with $\Omega_\phi=0.1$, 
$\Omega_c=0.85$ and $\Omega_b=0.05$ and the fractional energy denisty
in two species of massive neutrinos (solid line) 
with $\Omega_\nu=0.1$ (massive), 
$\Omega_c=0.85$ and $\Omega_b=0.05$.}\label{omegas}
\end{figure}
By analysing the simplified system we have been able to  get an idea
of what to expect from solving the full set of perturbation equations.
We found that the presence of $\phi$ in the matter era suppressed
the growth of perturbations by an easily determined factor.
The behaviour of this system is very much
like that of an MDM cosmology.
There one has, in addition to pressureless matter,
radiation, baryons and massless neutrinos, a component of matter
in two species of massive neutrinos with masses, $m_\nu$ of order a few eV.
The evolution of the energy density is similar to that of the
scalar field in the present case. Deep in the radiation era, 
these neutrinos behave as
a massless species and therefore the energy density scales as
radiation ($\propto 1/a^4$), while deep in the matter era they behave as 
non- relativistic matter which scales as $1/a^3$.  Unlike the scalar field
case, where the transition between the two regimes is 
determined by the transition from radiation to matter domination
the transition in the case of the neutrinos occurs when 
$3k_BT_\nu\simeq m_\nu$, where $k_B$ is 
the Boltzmann constant and $T\nu$ is the massive neutrino temperature.
This corresponds to redshift  
$z\simeq 1.8 \times 10^{5}\frac{m_\nu}{30 eV}$. 
 In Fig 
\ref{omegas} we plot the evolution of the energy densities in the
extra component in each of the two cosmologies. The 
energy density in the scalar field $\phi$ follows the transition 
very tightly while the energy density in neutrinos becomes 
non-relativistic after radiation-matter equality. In the latter
case there is a short period of time after equality when the massive 
neutrinos contribute less to the energy density than they
do asymptotically, and as a result the pressureless matter 
clumps more strongly for this period.

\begin{figure}
\centerline{\psfig{file=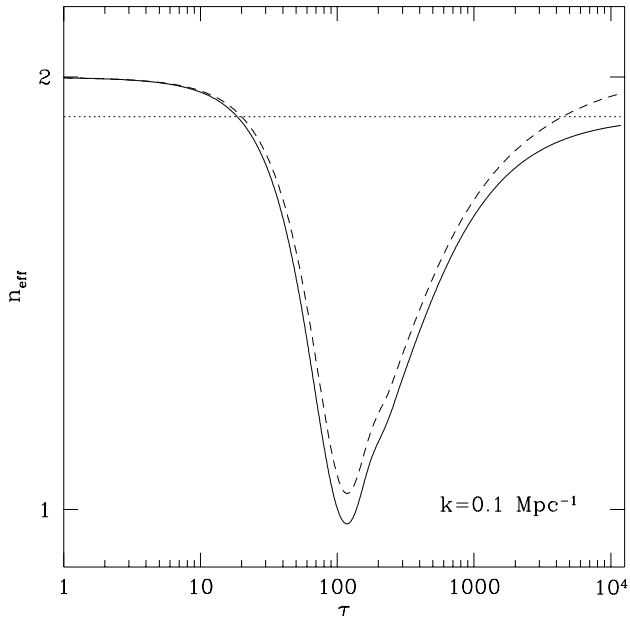,width=3.5in}}
\caption{We plot the evolution of $n_{eff}$ for $k=0.1$Mpc
in a MDM cosmology (dashed line) and in a universe
with $\Omega_\phi=0.1$ (solid line). The  dotted line
is the asymptotic matter era solution, $\epsilon$.
 }\label{neffnu.1}
\end{figure}

The correction to the exponent of the growing mode in the matter era
is also very similar in the two models. In  
\cite{BES} it was shown that the correction for MDM is 
$\epsilon_\nu=\frac{5}{2}(-1+
\sqrt{1-\frac{24}{25}\Omega_{\nu}})$, exactly the same as
we have just derived for the scalar field cosmology. The reason
is just that the sole assumption in the derivation of this 
result was that the exotic form of matter 
(the $\phi$ field in our case and the massive
neutrinos in the MDM case) does not cluster below a certain scale.
There is however again a small but important difference. For $\varphi$, 
the scale below which a given mode does not cluster is the horizon,
i.e. $\propto \frac{1}{\tau}$. Once it comes into the 
horizon it {\it never} clusters. For the massive neutrino
perturbations this is not the case. The free-streaming scale, 
when the neutrinos are non-relativistic is 
$k_{fs}=8a^{1/2}(m_\nu/10 eV)h$Mpc$^{-1}$,
which {\it grows} with time. Perturbations of a given wave number
 $k$ are damped while $k>k_{fs}$, but as soon as $k_{fs}$ becomes
small enough they behave like perturbations in pressureless matter 
and grow. So perturbations in an intermediate range of wavenumbers, 
despite suppression during a finite period of time, catch up on
the pressureless matter perturbations and contribute to gravitational 
collapse. This effect can be seen when we compare $n_{eff}$ for 
both these scenarios for $k=1$Mpc (i.e. a mode that comes into
the horizon at radiation-matter equality). In Fig \ref{neffnu.1} we
see that, while in the $\phi$CDM case $n_{eff}\rightarrow \epsilon$,
in the MDM case, perturbations in massive neutrinos start to grow
as pressureless matter and $n_{eff}\rightarrow 2$.

\begin{figure}
\centerline{\psfig{file=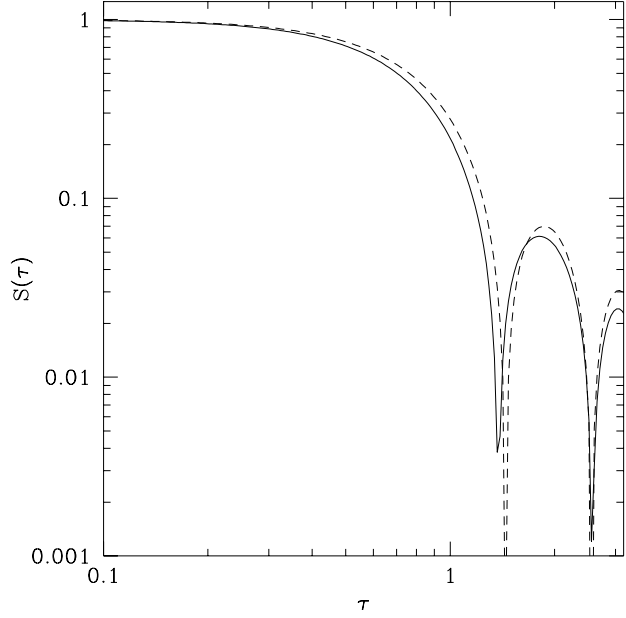,width=3.5in}}
\caption{The evolution of ${\cal S}$ at horizon crossing
($k=5$hMpc$^{-1}$) for the scalar field cosmology (solid line)
and for the mixed dark matter cosmology (dashed line). 
 }\label{sourcerad}
\end{figure}

A further subtle difference between the two scenarios can be seen
at the transition between super- and subhorizon behaviour in the 
radiation era. As shown above, the subhorizon
evolution of $\delta_c$ is $\propto$ constant, $\ln \tau$. The initial 
amplitude of these solutions is set by the amplitude at horizon
crossing, i.e. when $k\tau\simeq 1$. In the MDM case, the analysis
is also simple. The neutrinos are effectively massless and
therefore behave exactly like radiation. So again the subhorizon
evolution of $\delta_c$ is $\propto$ constant, $\ln \tau$. However
the transition to subhorizon evolution happens at smaller scales
than for the $\varphi$. Indeed from the equations we
see that the transition should happen when $c_sk\tau\simeq 1$ (where
$c^2_s=\frac{1}{3}$). A simple way of seeing this is by
looking at the source term for the $\delta_c$ in the radiation era,
$4{\cal S}(k,\tau)={\ddot \delta}_c+\frac{1}{\tau}{\dot \delta}_c
=\partial_\tau(\tau n_{eff})\frac{\delta_c}{\tau^2}$. On superhorizon
scales ${\cal S}(k,\tau)=1$, and 
the time when this quantity starts to deviate from a constant 
indicates the transition from
super- to subhorizon behaviour.
We have
\begin{eqnarray}
{\cal S}(k,\tau)=\frac{3}{4}{\cal H}^2(\Omega_\gamma\delta_\gamma
+\Omega_\nu\delta_\nu)
\end{eqnarray}
in the MDM scenario where we have grouped the massless and massive
neutrinos together, and we have
\begin{eqnarray}
{\cal S}(k,\tau)=\frac{3}{4}{\cal H}^2(\Omega_\gamma\delta_\gamma
+\Omega_\nu\delta_\nu)+\frac{2}{\lambda \tau}{\dot \varphi}+
\frac{1}{2\tau^2}\varphi
\end{eqnarray}
in the $\phi CDM$ case. In Fig. \ref{sourcerad} we look at the time evolution
of a mode with $k=5$h Mpc$^{-1}$ and see that transition to subhorizon
evolution occurs earlier in the scalar field scenario than in the
MDM scenario. As a result $\delta_c$ stops growing earlier
and there is an additional small, but non-negligable
suppression factor in the radiation era. 

\begin{figure}
\centerline{\psfig{file=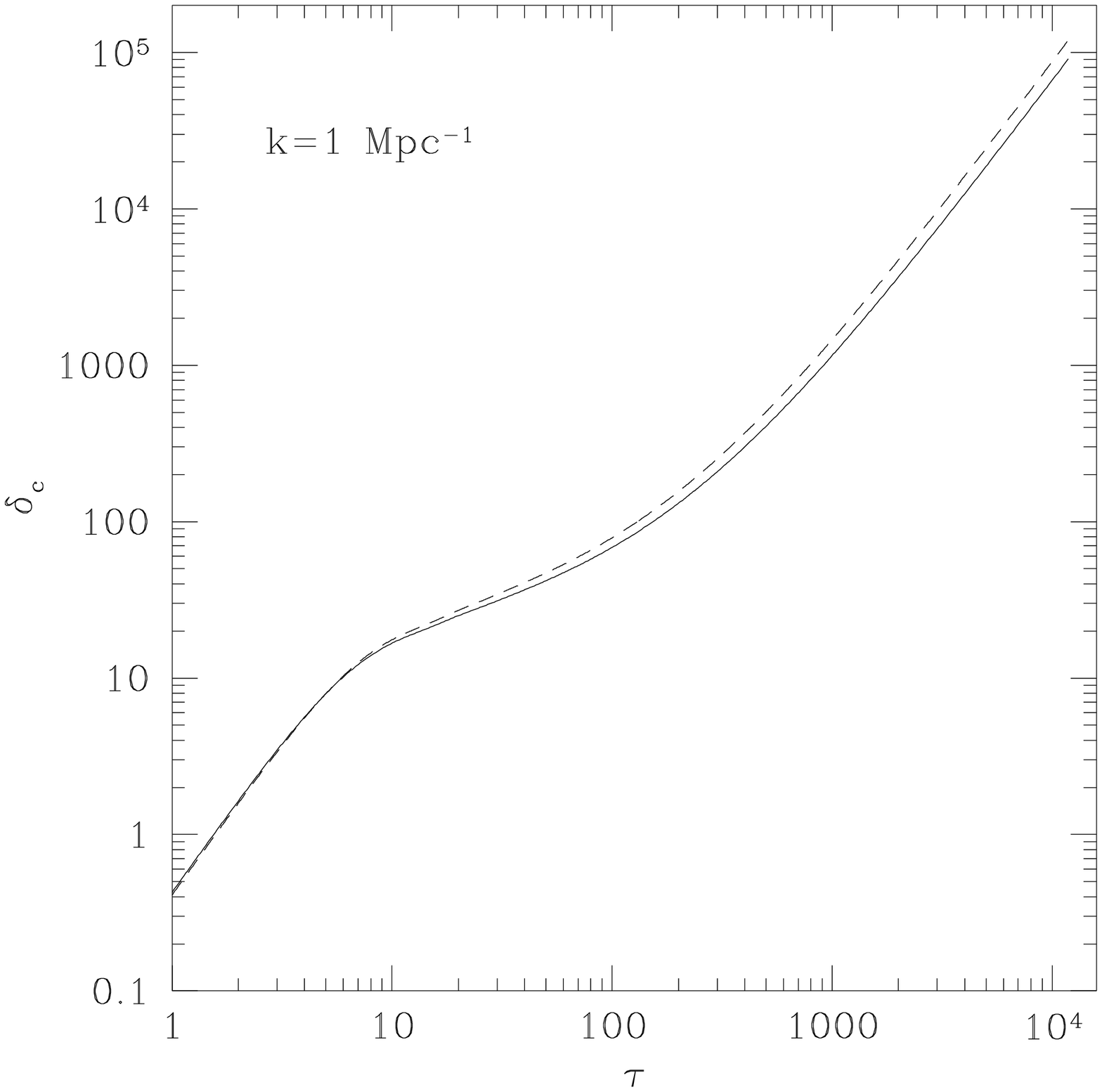,width=2.5in}
\psfig{file=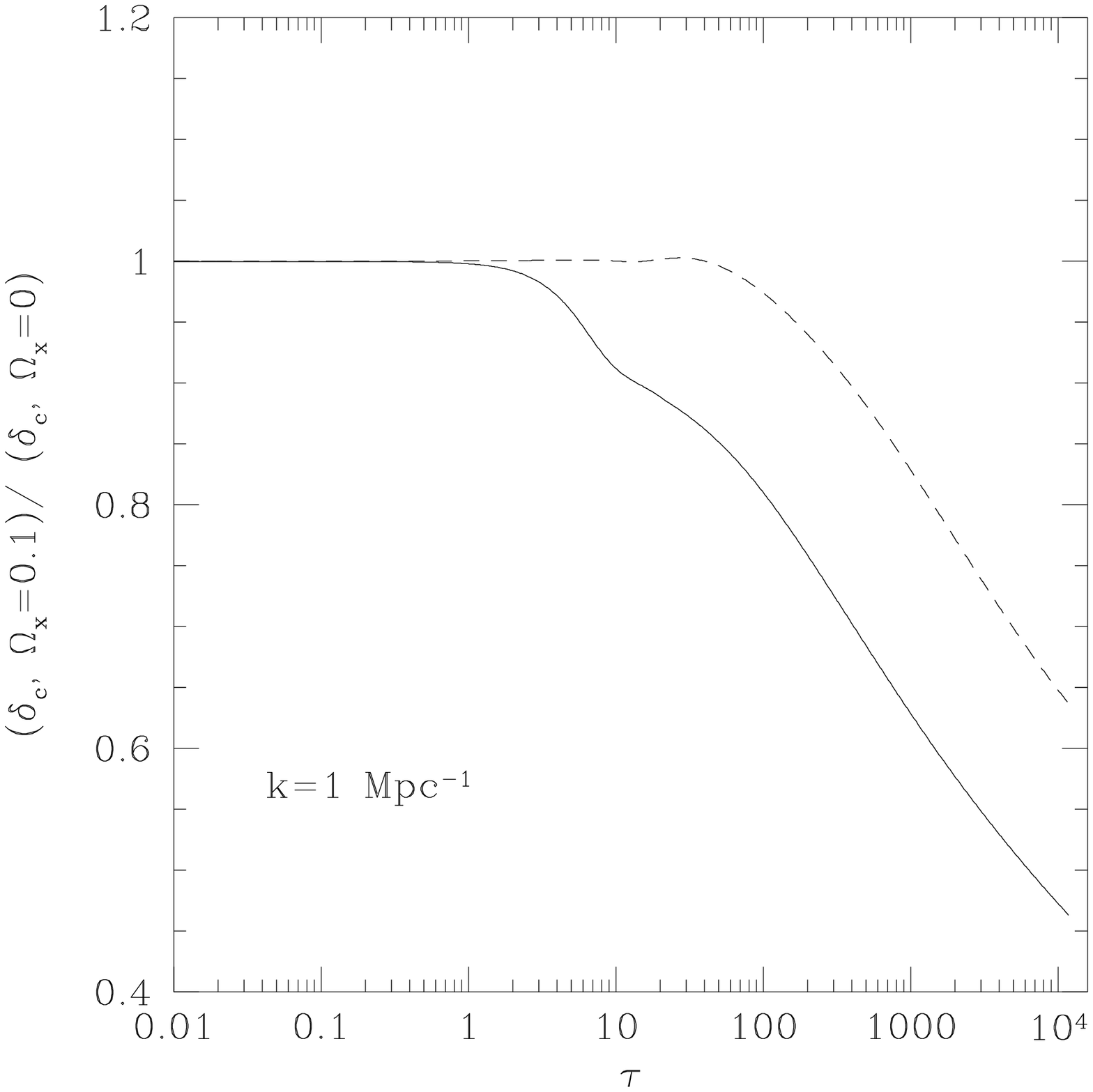,width=2.5in}\psfig{file=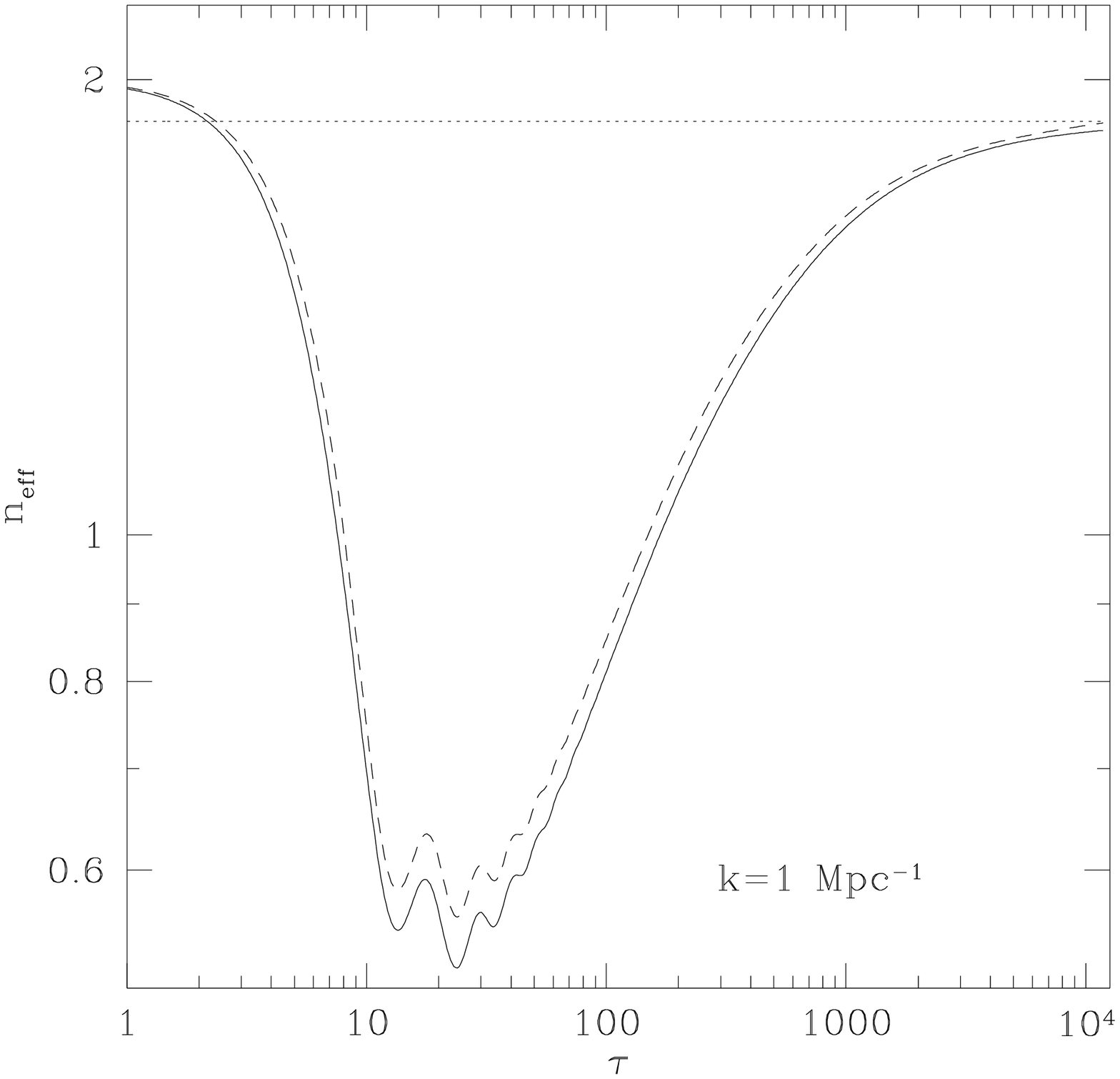,width=2.5in}}
\caption{In the left panel we plot the evolution 
of $\delta_c$ ($k=1$Mpc$^{-1}$) for
the scalar field cosmology (solid line) and the mixed dark matter
cosmology. In the center panel we plot the suppression factor relative
to the standard cold dark matter scenario. In the right panel
we plot the dimensionless growth rate for these two modes.}\label{hifreq}
\end{figure}

In Fig. \ref{hifreq}  we
compare the evolution of $\delta_c$ for these two models. It
is evident that the suppression of $\delta_c$ in the scalr field
model starts much earlier than in the MDM model. This is also
clear when we compare  $n_{eff}$ in the two  cases.

In conclusion, all these effects we have just discussed combine 
to give additional suppression in the $\phi$CDM model as compared to 
the MDM model. 

\subsection{The temperature anisotropy angular power
spectrum\label{Cell}}
To complete our analysis of the evolution of perturbations in
this cosmology, we shall now look at the effect of the scalar field
on the cosmic microwave background.
The temperature anisotropy measured in a given
direction of the sky can be expanded in spherical harmonics as
\begin{eqnarray}
\frac{\Delta T}{T}({\bf n})=\sum_{lm}a_{\ell m}Y_{\ell m}
({\bf n}) \label{almdef}
\end{eqnarray}
We work again in the framework of inflationary cosmologies so
the $a_{lm}$s are Gaussian random variables which satisfy
\begin{eqnarray}
\langle a^*_{\ell'm'}a_{\ell m}\rangle=C_\ell \delta_{\ell'\ell}\delta_{m'm} 
\end{eqnarray}
The angular power spectrum, $C_\ell$, contains all the
information about the statistical properties of the cosmic microwave
background. One can relate it to the temperature brightness function
we derived above through
\begin{eqnarray}
C_\ell=4\pi\int\frac{dk}{k}|\Delta_{T_\ell}(k,\tau_0)|^2
\end{eqnarray}
where we assume again  that we are considering the scale-invariant,
inflationary scenario.

In ref \cite{HS}, the authors presented a useful simplification
with which one can understand the various features in the
angular power spectrum. We shall present a simplified version,
in the synchronous gauge and use it to understand the effect 
$\Omega_\phi$ will have on the $C_\ell$s. To a reasonable
approximation we can write (defining $\chi=\frac{{\dot h}+6{\dot \eta}}{2k^2}$)
\begin{eqnarray}
\Delta_{T\ell}(k,\tau_0)&=&[\frac{1}{4}\delta_\gamma+2{\dot
\chi}](k,\tau_*)j_\ell(k(\tau_0-\tau_*))\nonumber \\ & &+
[\frac{\theta_b}{k^2}+ \chi][\frac{\ell}{2\ell+1} j_{\ell-1}
(k(\tau_0-\tau_*))-\frac{\ell+1}{2\ell+1}j_{\ell-1}
(k(\tau_0-\tau_*))] \nonumber \\ & &+\int_{\tau_*}^{\tau_0}({\dot \eta}(k,\tau)+{\ddot
\chi}(k,\tau))j_\ell(k(\tau_0-\tau))\label{cmbana}
\end{eqnarray}
assuming instantaneous recombination at $\tau_*$ (we
do not consider Silk damping in this discussion). 

\begin{figure}
\centerline{\psfig{file=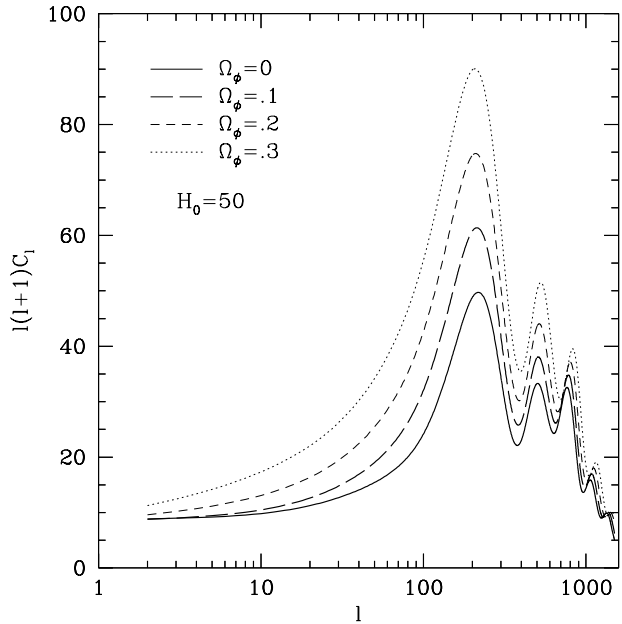,width=3.5in}}
\caption{The angular power spectrum of CMB anisotropies for
different values of $\Omega_\phi$. These power spectra are not
normalised to COBE.}\label{celles}
\end{figure}
\begin{figure}
\centerline{\psfig{file=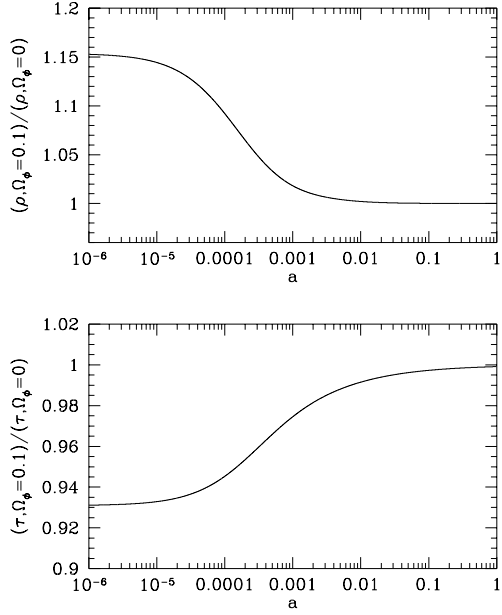,width=3.5in}}
\caption{The top panel shows the ratio 
of the total energy density in a universe
with $\Omega_\phi=0.1$ to that in a scalar field free universe.
The bottom panel 
shows the ratio of conformal horizons.}\label{sound}
\end{figure}
\begin{figure}
\centerline{\psfig{file=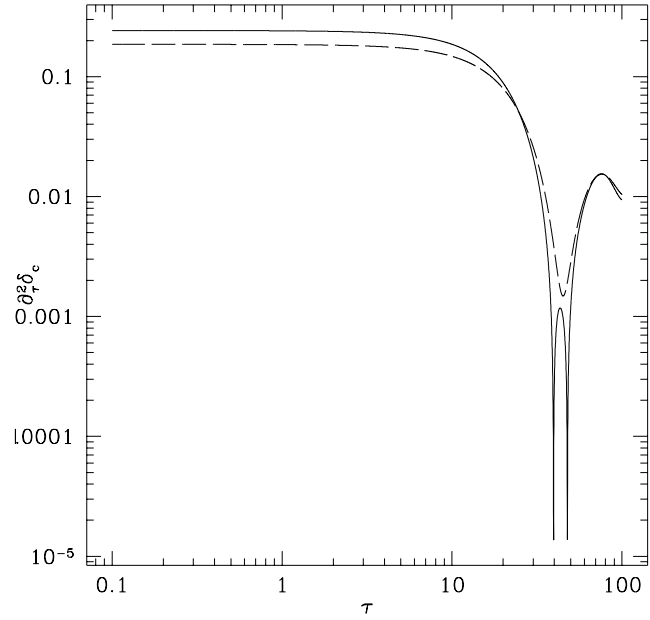,width=3.5in}}
\caption{A comparison of the source of acoustic oscillations in
$\delta_\gamma$ for $k=0.2$hMpc$^{-1}$ for SCDM (dashed) and $\phi$CDM
(solid)
with
$\Omega_\phi=0.1$ and $H_0=50$Kms$^{-1}$Mpc$^{-1}$.}\label{ddotc}
\end{figure}
In Fig \ref{celles} we plot the angular power spectrum of
the CMB for a few values of $C_\ell$. Comparing to a SCDM
$C_\ell$ we see two effects. Firstly, for $\ell>200$
we see that the peaks are slightly shifted to smaller scales.
In \cite{DGS} a clear analysis was presented which explains the location
of the peaks in the MDM model scenario, and we will now see that
the same effect is present here, {\it albeit in the opposite direction}. 
The anisotropies at these scales are generated primarily at $\tau_*$ and 
for the purpose of this discussion we can assume that the first 
term in (\ref{cmbana}) dominates.
As shown in \cite{HS}
the peak structure of this term is given by
\begin{eqnarray}
\frac{1}{4}\delta_\gamma+2{\dot\chi}\propto \cos k r_s(\tau_*)
\end{eqnarray}
where $r_s$ is the sound horizon in the baryon-photon fluid,
$r_s(\tau)=\int_0^\tau \frac{d\tau'}{3[1+R(\tau')]}$. Now one of
the effects described in \cite{DGS} and which we see here is that
for a given redshift (during a period of time which contains 
the time of recombination), $r_s$  is different in the
presence of the scalar field than in it absence. This is to be expected
as deep in the radiation era we now have three components
contributing, the radiation (whose energy density is set by $T_{CMB}$,
the massless neutrinos (whose energy density is set by $T_{\nu}$) and
the scalar field whose energy density is a constant fraction
of the total energy density. Therefore, for a given redshift,
the total energy density in the presence of the scalar field
is larger than the total energy density in the absence of the
scalar field. As a consequence the expansion rate will be larger
and the conformal horizon will be smaller. In Fig. \ref{sound} 
we plot the ratio of energy densities and the ratio of comoving
horizons as a function of scale factor. At recombination
($a\simeq10^{-3}$), the
conformal horizon (and consequently the sound horizon) is smaller,
and as seen in Fig. \ref{celles} the peaks are shifted to the right.

Finally, the other effect $\Omega_\phi$ has on the peaks is to boost
their amplitude.  Let us assume
$c_s^2=\frac{1}{3}$, i.e. we are deep in the radiation era.
As shown in the section \ref{pertdens} we can write
\begin{eqnarray}
{\ddot \delta}_\gamma+\frac{k^2}{3}\delta_\gamma=\frac{4}{3}{\ddot \delta}_c
\end{eqnarray}
so the acoustic oscillations are sourced by ${\ddot \delta}_c$. One
finds that this source is larger in the presence
of the scalar field than in its absence (see Fig. \ref{ddotc}).
The increase in the amplitude of this driving term will increase
the amplitude in $\delta_\gamma$ and lead to the increase by a few
percent of the acoustic peaks of the angular power spectrum.

\section{Constraints from COBE and large scale structure}\label{data}

In the previous section we have discussed in some detail the
evolution of density perturbations and temperature anisotropies
in the $\phi$CDM cosmology. This approximate analysis indicated
that results should be similar to those in the MDM model, except that
there should be additional suppression of power in $\Delta^2(k)$
on small scales. Using the full results of our numerical evolution
we  now compare our model with the observational constraints which 
measure fluctuations on a wide range of scales.
We first compare  the $C_\ell$s to the
COBE data and calculate the normalization of
the perturbations model for different $\Omega_\phi$. 
For completeness we plot a selection of $C_\ell$ compared to a 
compilation of data sets and other candidate theories. 
We then compare  $\Delta^2(k)$ with the
observational $\Delta^2(k)$ rendered in \cite{PD} from a
compilation of surveys. For this section we 
define $\phi$CDM$_1$ to be a universe with $\Omega_\phi=0.08$,
$H_0=50$kms$^{-1}$Mpc$^{-1}$ and $\phi$CDM$_2$ to be a universe
with $\Omega_\phi=0.12$, $H_0=65$kms$^{-1}$Mpc$^{-1}$.

\begin{figure}
\centerline{\psfig{file=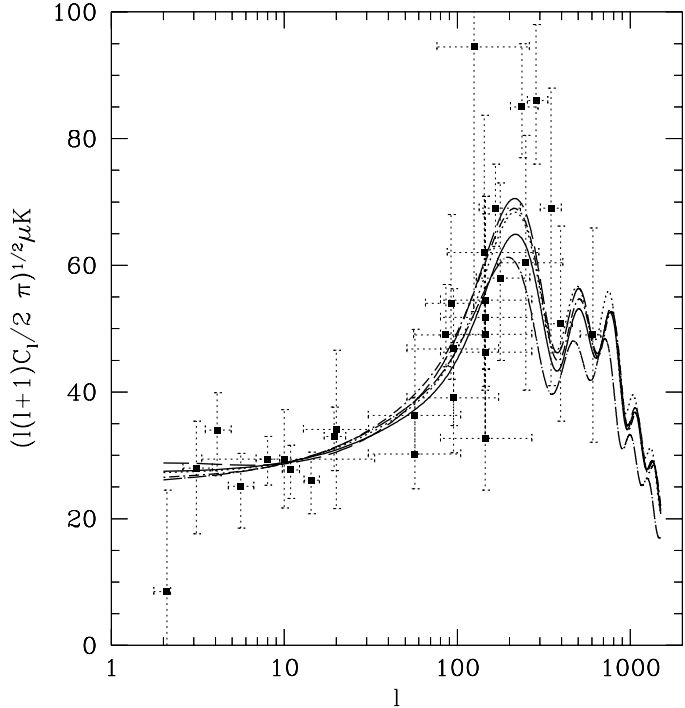,width=4in}}
\caption{A comparison of the angular power spectra of temperature
anisotropies for five
COBE normalized models with the current experimental situation.
The models are sCDM in solid, 
($h=.5$), $\Lambda$CDM in long dash, ($\Omega_\Lambda=.6$, $h=.65$),
MDM in dotted,  ($\Omega_\nu=.2$ and $h=.5$), $\phi$CDM in
dot-short dash,  ($\Omega_\phi=.08$, $h=.5$) and $\phi$CDM in
dot-long dash,  ($\Omega_\phi=.12$, $h=.65$).
All of them have $\Omega_bh^2=0.0125$.}\label{dtfull}
\end{figure}

The past five years has seen a tremendous growth in
experimental physics of the CMB. Over twenty
experimental groups have reported detections of fluctuations
in the CMB and a rough picture is emerging of 
the angular power spectrum. It is fair to say that the
most uncontroversial and useful measurement that we have
is that of  COBE, which  tells us that on scales
larger than $10^\circ$ the fluctuation are approximately scale invariant
with a $Q_{rms}=18\mu K$. Measurements on smaller scales
seem to indicate a rise in the power spectrum, but
a convincing constraint is still lacking.  In Fig \ref{dtfull}
we present a compilation of measurements of \cite{TH} as compared to
two $\phi$CDM models and a few candidate rival models.
Clearly there is still is a large spread although an
overall shape is emerging. 

\begin{figure}
\centerline{\psfig{file=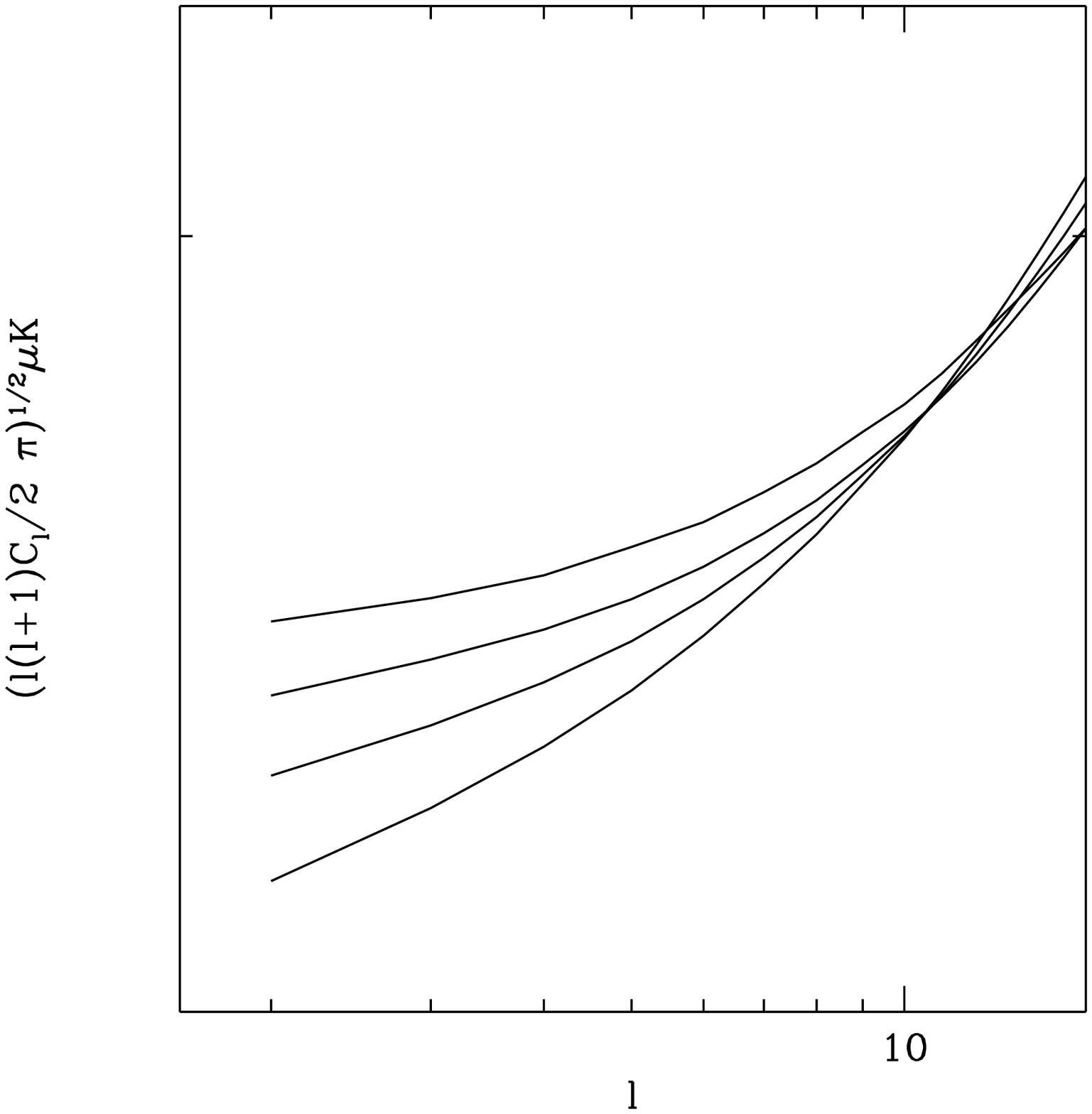,width=3.5in}
\psfig{file=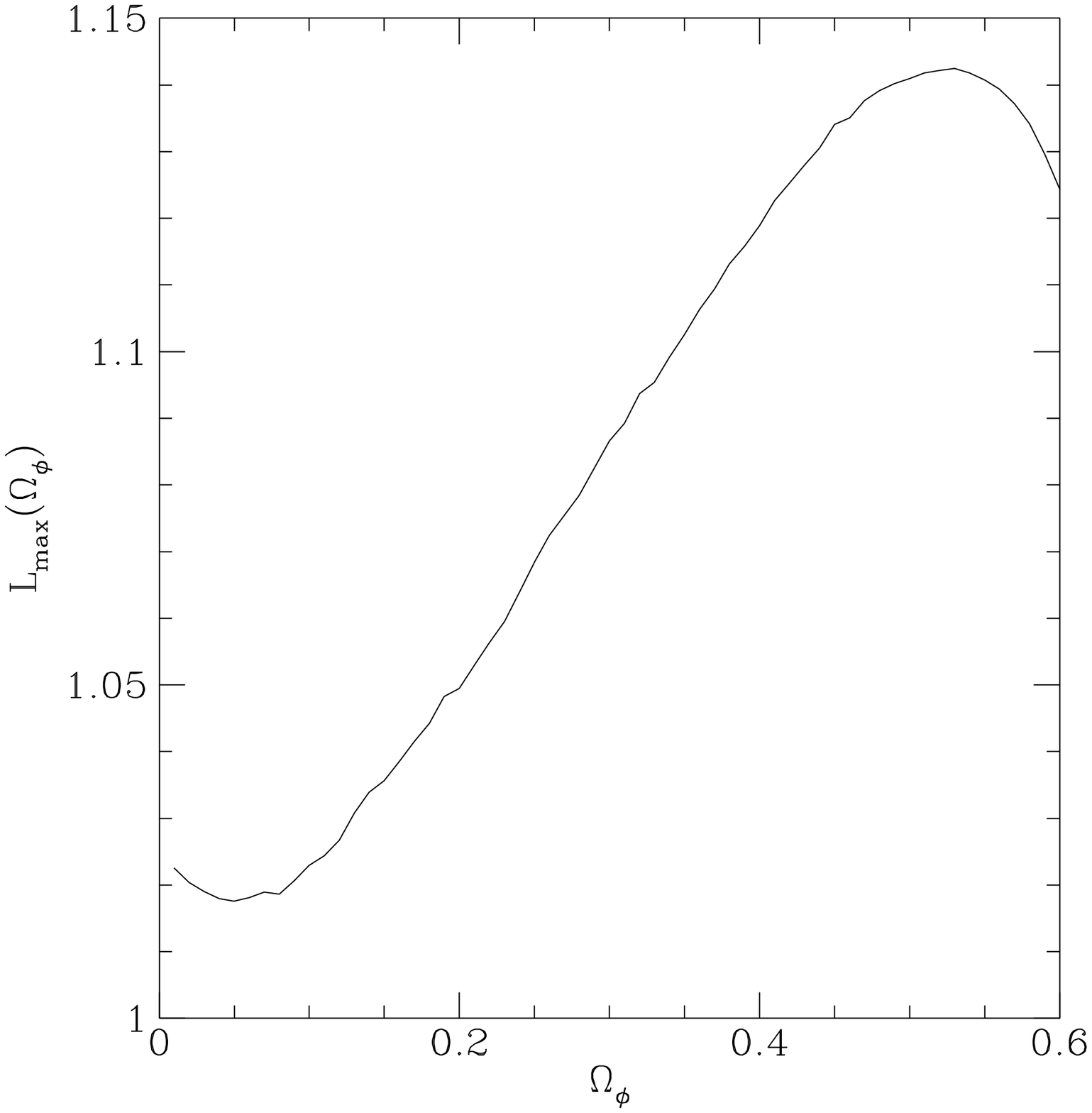,width=3.5in}}
\caption{The left panel shows the small $\ell$ angular power spectrum
for a family of $\phi$CDM models. The models are $\Omega_\phi$= 0,
0.04, 0.08 and 0.12 in order of increasing ``tilt''. The right
panel shows the maximum likelihood of the best fit model as a
function of $\Omega_\phi$.
 We have fixed $h=0.5$ and $\Omega_bh^2=0.0125$.}\label{larga}
\end{figure}

In the previous section we described the effect that $\Omega_\phi$
would have on the $C_\ell$s. For small $\ell$s the dominant
effect to note is the increase in power of the acoustic peaks
relative to the large-scale, scale-invariant plateau. The
larger ia $\Omega_\phi$, the larger is the boost and therefore
the smaller the $\ell$s which are affected. In practice it
introduces an effective ``tilt'' in the large angle power
spectrum as can be seen in Fig \ref{larga}. 
It is useful to quantify how good a fit $\phi$CDM $C_\ell$s are
to the COBE data. The correct framework to work with is
maximum likelihood analysis. In \cite{BW} the authors have
supplied us with an efficient way of evaluating the
likelihood of a given mode relative to purely scale invariant
fluctuations. On the left panel we plot the likelihood of the
best fit model as a function of $\Omega_\phi$. There are
two important things to note. Firstly the well known fact
that, if one includes the quadrupole, the COBE data 
favours more tilted models and therefore a larger
$\Omega_\phi$. Secondly, the likelihood function for
$\Omega_\phi$ should be very flat; indeed the dependence of the
tilt on $\Omega_\phi$ is very weak. As explained in the previous 
section the effect of $\Omega_\phi$ on the $C_\ell$s is of
order a few percent and concentrated at large $\ell$s. There
will be little variation on COBE scales.

One of the key observational constraints for any class of models
is the mass variance per unit interval in $\ln k$ as defined in 
Eq. \ref{delta}. This quantifies the amount of clustering over a
range of scales. In \cite{PD} the authors compiled a series of surveys and
attempted to extract what they believe to be the underlying
$\Delta^2(k)$ of the linear density field. This involved
a series of corrections: Firstly, the assumption that
the different samples were biased in different ways with
respect to the underlying density field; secondly, that there
are redshift distortions in the observed structures; and
finally, that some of the structures have undergone non-linear
collapse. This final correction is model dependent and
in principal great care should be taken in making definitive
comparisons between our theoretical $\Delta^2(k)$ and that
presented in \cite{PD}. In practice we shall assume that
possible corrections are small and compare them.
In a future publication we shall analyse the non-linear features
of this theory.

\begin{figure}
\centerline{\psfig{file=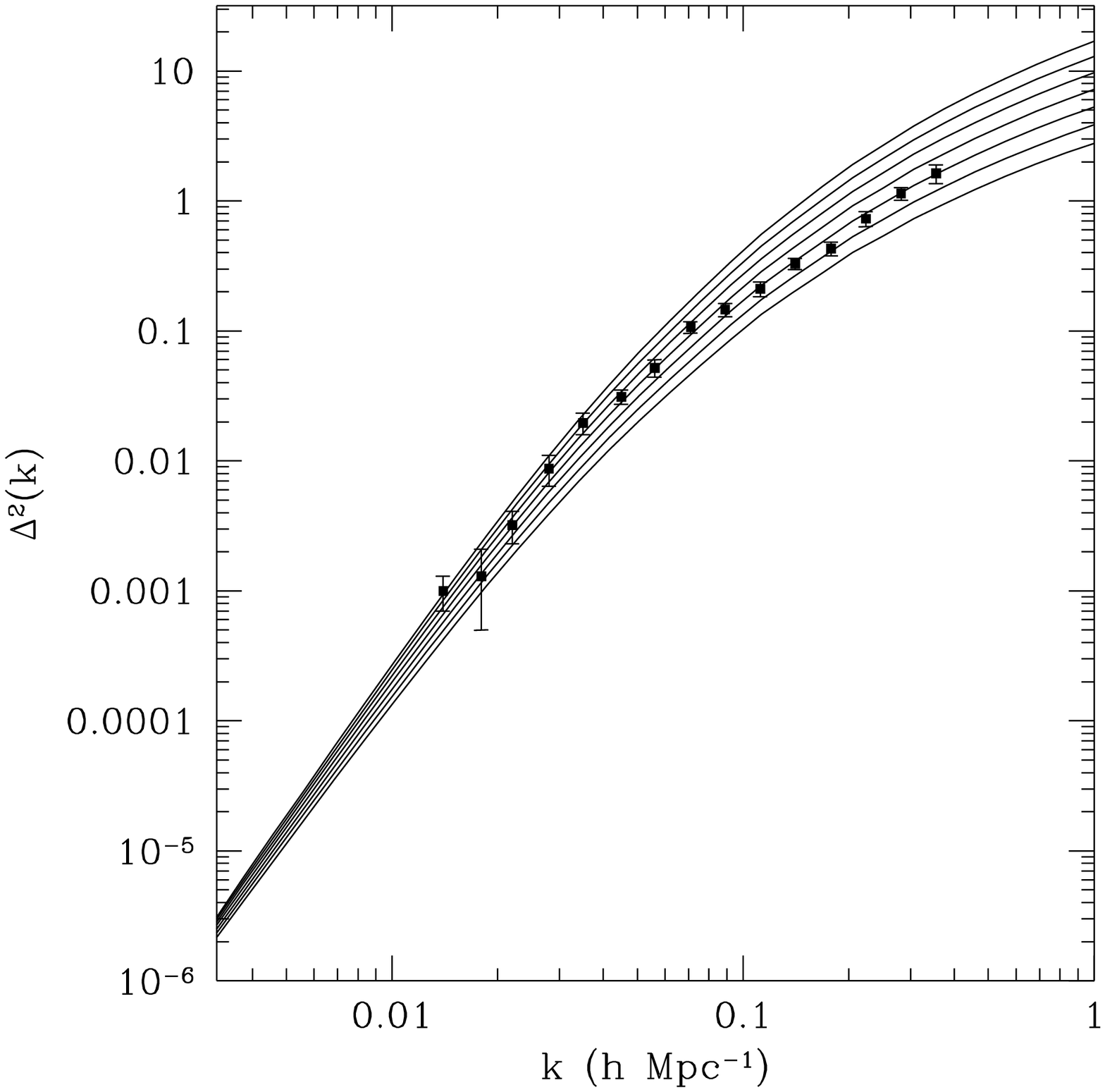,width=3.5in}
\psfig{file=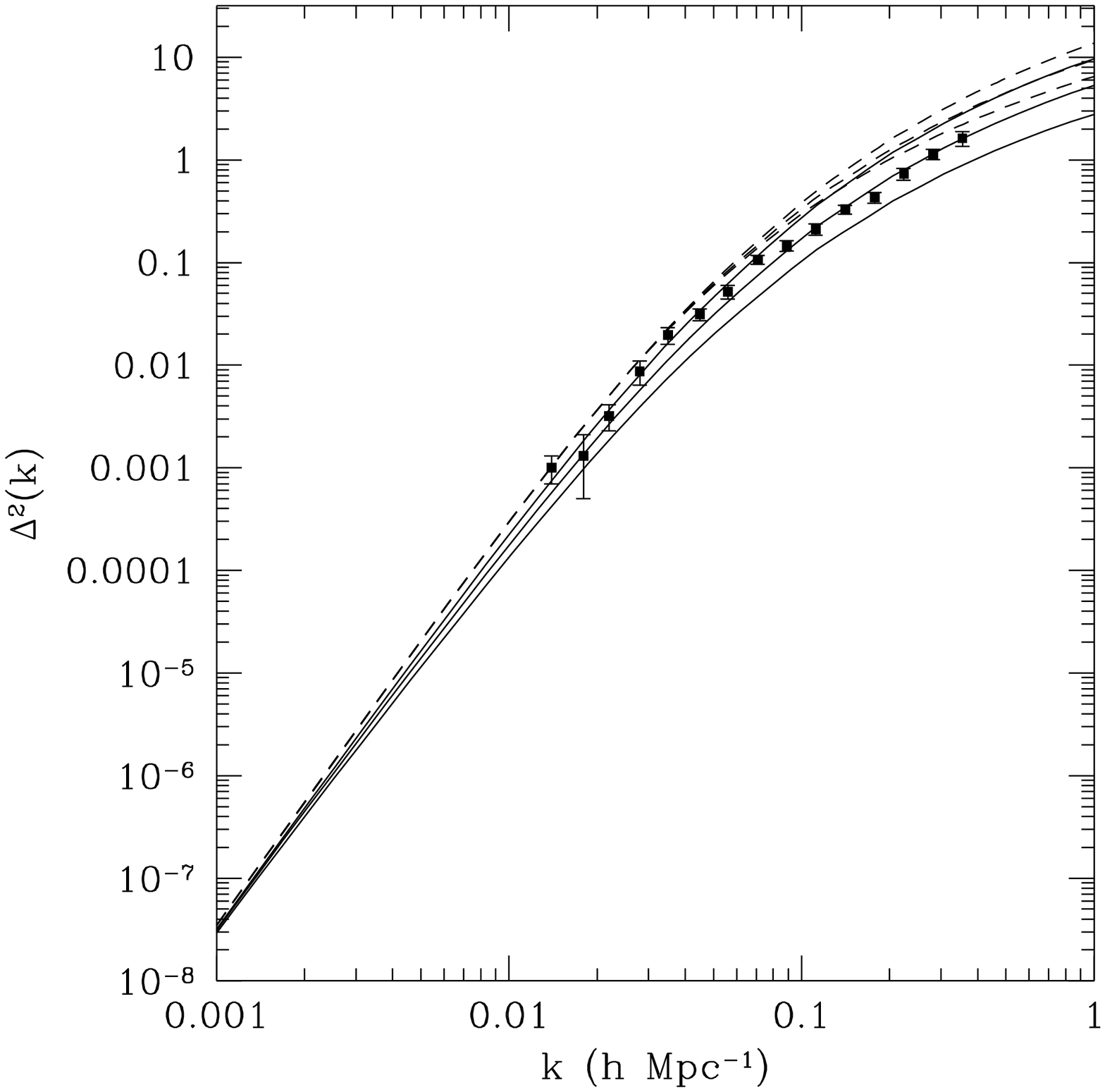,width=3.5in}}
\caption{On the left panel we plot the dependence of 
$\Delta^2(k)=\frac{d\sigma^2}{d \ln k}$
on $\Omega_\phi$. For $h=.5$, $\Omega_bh^2=0.0125$ we show
(in order of decreasing amplitude) plots for $\Omega_\phi=0$, $.02$,
$.04$, $.06$, $.08$, $.10$ and $.12$. In the right panel we compare a family
of MDM models (dashed, with $\Omega_\nu=0.04$, $0.08$ and $0.12$ in 
order of decreasing amplitude) with a family of $\phi$CDM models 
(dashed, with $\Omega_\phi=0.04$, $0.08$ and $0.12$ in 
order of decreasing amplitude).
}\label{deltaphis}
\end{figure}
\begin{figure}
\centerline{\psfig{file=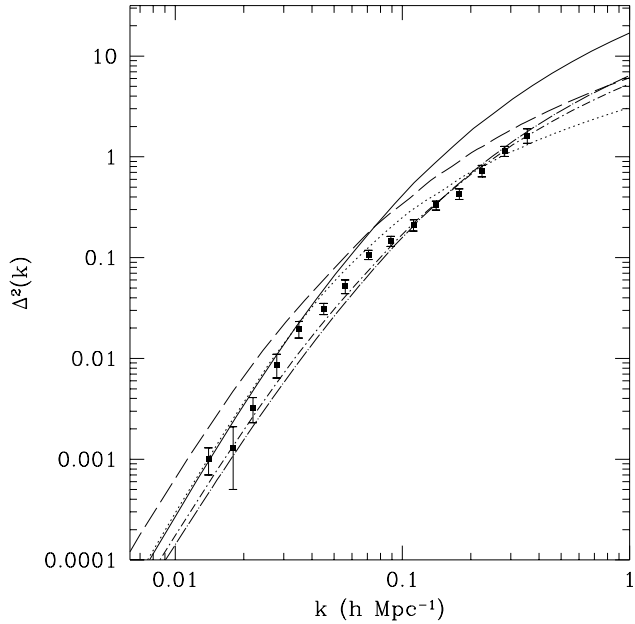,width=3.5in}}
\caption{A
 comparison of the  mass variance per unit interval in $\ln
k$, $\Delta^2(k)=\frac{d\sigma^2}{d \ln k}$,
for five COBE normalized models 
with a rendition of the linear power spectrum from
various data sets (corrections for non-linearity, redshift distortions
and biasing have been introduced ). 
The models are sCDM in solid
($h=.5$), $\Lambda$CDM in long dash ($\Omega_\Lambda=.6$, $h=.65$),
MDM in dotted ($\Omega_\nu=.2$ and $h=.5$), $\phi$CDM in
dot-short dash ($\Omega_\phi=.08$, $h=.5$) and $\phi$CDM in
dot-long dash ($\Omega_\phi=.12$, $h=.65$).
All of them have $\Omega_bh^2=0.0125$.
}\label{PDfig}
\end{figure}

In Fig. \ref{deltaphis} we plot a family of COBE normalized 
$\Delta^2(k)$ with $h=0.5$ and $\Omega_bh^2=0.0125$. We clearly see
the features described in \ref{eqpert}, i.e. the larger the
$\Omega_\phi$, the smaller the $k$ for which $\Delta^2(k)$ departs
from scale invariance. In the other panel we can also see how
it differs from a MDM model with the same background cosmological
parameters. For the same energy density in exotic matter component
(i.e. $\phi$ or massive $\nu$) there is more suppression in
the $\phi$CDM case. Finally we see how it compares to the
data of \cite{PD}. We find that, for 
$\Omega_\phi$ in the range $0.08-0.12$ 
we can match the data
with as good agreement as the MDM model and some other candidate
models. This is displayed in Fig \ref{PDfig}. In the following table 
 we tabulate the $\chi_2$ values (with 15 degrees of
freedom) of these models, in increasing goodness of fit.
$\phi CDM$ is competetive with the best fit model of MDM.
\begin{center}
\label{table1}
\begin{tabular}{|c|c|}
\hline SCDM & 103.96 \\
\hline $\Lambda$CDM & 52.5 \\
\hline $\phi$CDM$_2$ & 14.5 \\
\hline MDM & 10.25 \\
\hline $\phi$CDM$_1$ & 7.53 \\
\hline 
\end{tabular}
\end{center}
One can fit $\Delta^2(k)$ for these models (to 10$\%$ 
in the range $0\le \Omega_\phi \le 0.16$) with:
\begin{eqnarray}
\Delta^2(k)&=&D(k,\Omega_\phi)\Delta^2(k)_{CDM} \nonumber \\
D(k,\Omega_\phi)&=&(1+1.5\Omega_\phi-10\Omega_\phi^2)\left (
\frac{1+5k^{1.1295}+4k^{2.259}}{1+1\times10^7k^{2.259}} \right )^{1.15\Omega_\phi^{1.01}}
\end{eqnarray}
where $\Delta^2(k)_{CDM}$ is  the COBE normalized CDM mass variance,
and contains all the dependence on remaining cosmological parameters
such as $H_0$ and $\Omega_b$ \cite{commBBKS}.

A useful quantity to work with is that characterising
the mass fluctuations on $R$h$^{-1}$Mpc scales
\begin{eqnarray}
\sigma^2(R)=\int_0^\infty\frac{dk}{k}\Delta^2(k)
(\frac{3j_1(kR)}{kR})^2.
\end{eqnarray}
In particular it has become the norm to compare this quantity
at $8$h$^{-1}$Mpc with the abundances of rich clusters. 
It is premature to use this constraint
with the our current understanding of $\phi$CDM. The best
measurements of such abundances involve an estimate of
the number density of X ray clusters of a given surface temperature.
To relate these temperatures to masses in an accurate way one
has to rely on N-body simulations of clusters. This has been done
for a few cosmologies and we can use the results they use as
a rough guide but care should be taken with using such results
at face value. The fact that we can fit $\Delta^2(k)$ to the
data of \cite{PD} is already strong indication that we are on
the right track. If we use the values of \cite{WEF} we have
$\sigma_8\simeq0.5-0.8$. A good fit to $\sigma_8$ is
\begin{eqnarray}
\sigma_8(\Omega_\phi)=e^{-8.7\Omega_\phi^{1.15}}\sigma^{CDM}_8
\end{eqnarray}
where $\sigma^{CDM}_8$ is the COBE normalized sCDM $\sigma_8$.
As would be expected, for the range of values for which we get a good
agreement with \cite{PD}, we also match the cluster abundance
constraints.

Finally it is desirable to make a comparison with some measure of small
scale clustering at early times. In \cite{LLSSV}, the authors
used a simple analytic estimate of the fraction of collapsed
objects at redshift $z=3$ and $z=4$ to
show that for $\Omega_\nu=0.3$, MDM models predict too little
structure as compared to that inferred from the Lyman-$\alpha$
measurements \cite{LS}. More recently in \cite{he}, the authors
considered a larger range of cosmological parameters and found
that constraints from the Lyman-$\alpha$ systems could be sufficiently
restrictive to rule out a large range of  models. From
Fig \ref{PDfig} we can see that $\phi$CDM should fare better than
MDM on very small scales. This is easy to understand: We argued in
Section
\ref{pert} that effectiveness of $\phi$CDM was mainly due to the fact
that the scalar field free-streaming scale grows with time while
the massive neutrino free streaming scale decays with time. 
We then need a larger amount of massive neutrinos to fit
both COBE and the cluster abundances in the MDM model than
the amount of scalar field in $\phi$CDM. On much smaller scales
(the scales probed by Lyman-$\alpha$ systems) i.e. scales smaller
than the massive neutrino free streaming scale, perturbations
in MDM should be more supressed than in $\phi$CDM. This means
$\phi$CDM should fare better than MDM with regards to the
Lyman-$\alpha$ constraints. A
preliminary check on our model can be done using the technique of
\cite{LLSSV}. In brief one can make a conservative estimate that a
fraction $f_{gas}$ of matter is in gas at that time and set
bounds on the amount of objects with masses greater than 
$10^{10} (1-\Omega_\phi)^{-1}h^{-1}M_{\odot}$.
Using the Press-Schecter
formalism one can then derive a bound:
\begin{eqnarray}
{\rm erfc}(\frac{1.7}{\sqrt{2}\sigma(R,z)})&>&0.16h/f_{gas} \ \ \ z=3
\nonumber \\
{\rm erfc}(\frac{1.7}{\sqrt{2}\sigma(R,z)})&>&0.104h/f_{gas} \ \ \ z=4
\end{eqnarray} 
where erfc is the complementary error function and $R=0.1-0.2$Mpc.
For $\phi$CDM in the range of $h$ and
$\Omega_\phi$ that we have been considering we find that
if  $f_{gas}=1$, these models are consistent with this constraint.
Note that the MDM model already has serious problems with this
conservative constraint.
If we consider a less a conservative constraint and take
$f_{gas}\sim0.1$ as seems to be indicated by hydrodynamical studies,
than $\phi$CDM is inconsistent with these measurements. More detailed
observations and modeling of Lyman-$\alpha$ systems will supply us
with a very strong constraint on this class of models.

\begin{figure}
\centerline{\psfig{file=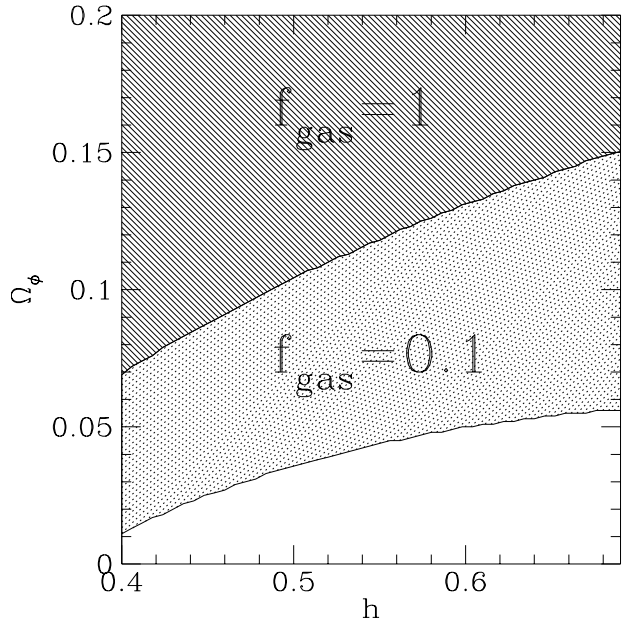,width=3.5in}}
\caption{The shaded region represents the allowed region of parameter
space consistent with the Lyman-$\alpha$ constraints of
\protect\cite{LS}. The solid hatched region corresponds to
the values of $h$ and $\Omega_\phi$ excluded for $f_{gas}=1$.
The dotted hatched region corresponds to the 
values of $h$ and $\Omega_\phi$ excluded for $f_{gas}=0.1$.
}\label{lyman}
\end{figure}

\section{Conclusions}\label{conc}

In the first part of this paper we gave the motivation for
considering the particular scalar field cosmology we have now studied in
detail: The addition made to the standard SCDM model does not
involve any tuning of the type involved in other modifications,
in that no energy scale characteristic of the universe at recent
epochs is invoked. The form of the required potential is one which 
arises in many particle physics models, with values of the single 
free parameter of the order required. Having analysed the model 
and determined the best fit to structure formation, we now first
comment on these aspects of the model. 
 
In Section \ref{ics} we argued that, in typical inflationary models
with the usual mechanism of reheating, one would expect the
attractor solution to be established well prior to nucleosynthesis.
In this case we should therefore compare our best fit to
the constraint from the latter, as given in 
(\ref{nucleo-constraint}), which is satisfied (clearly for
the conservative bound, and marginally for the tighter one).
Further, in this case we need to check that
the value $\Omega_b h^2=0.0125$ which  is equivalent to the baryon 
to photon ratio $\eta_{10}=3.3$ (in units of $10^{-10}$)
is within the allowed range at the corresponding expansion rate.
This is not a question which is easy to answer simply because
most nucleosynthesis calculations give results only in terms of
the range of $\eta_{10}$ allowed at the standard model  expansion 
rate, or the maximum additonal number degree of freedom allowed.
For the conservative criteria \cite{ks,sarkar-review} which gave 
the weaker upper bound in (\ref{nucleo-constraint})  the allowed 
range narrows from $\eta_{10} \in [1.65, 8.9]$ at $\Omega_\phi=0$ 
to the lower bound $\eta_{10}=1.65$ at $\Omega_\phi=0.15$.
Extrapolating the use of these criteria for a more
restrictive case for which the required data is given in \cite{bs}
(i.e. the allowed range of $\eta_{10}$ is given as a function of the 
energy density in an extra component) the value $\eta_{10}=3.3$ would
appear to be in the allowed range at $\Omega_\phi=0.1$. 
For a more restrictive set of nucleosynthesis constraints a slightly
lower value of $\Omega_bh^2$ might be required for consistency.
 
In an alternative model of reheating (which is by construction
associated with the existence of this same type of potential), 
we saw that the time of re-entry to the attractor could be  
after nucleosynthesis, at a time which depends on both 
$\Omega_\phi$ and the other parameter in this model $H_i$, 
the expansion rate at the end of inflation. 
By making the assumption that the attractor is established prior to
nucleosynthesis we restricted ourselves to a (large) part of 
parameter space.
From Fig \ref{HvsOmega} we can read off that this corresponds,
for $\Omega_\phi \approx 0.1$, to $H_i > 10^{14}$GeV
or $\rho_i^{1/4} > 10^{16}$ GeV. As we noted at the end of 
section this is consistent with what would guess would be the most
natural range of these parameters in this model. On the other hand,
it is inconsistent with models in which the phase of scalar field
domination continues until just before  nucleosynthesis (with 
the consequences described in \cite{mjtp}) since they correspond 
to the line defining the lower bound from nucleosynthesis  
in Fig \ref{HvsOmega}. For entry to the attractor before today,
it can be see from Fig \ref{HvsOmega} that one requires in this 
case $\Omega_\phi > 0.22$. It would be interesting to
analyse the effect on structure formation in these models,
and indeed in all of the parameter space for this model excluded by us 
in our present analysis.

In terms of $\lambda$ our best-fit corresponds to the range
$\lambda \in [5,6.1]$. As we discussed in Section \ref{exptls}
this value is certainly of the order observed in the fundamental
particle physical theories of which they have been observed to be
a generic feature,  and may even be in the precise range found in
certain theories. This suggests the exciting possibility of 
ultimately linking the cosmological features which would provide a 
signature for these fields to details of physics at the Planck scale. 
From the point of view of particle physics motivated model-building 
it would be particularly interesting also to look at models where the
simple exponential potential represents the asymptotic behaviour
of a potential which can support inflation in another region,
since this would be likely to produce a very constrained model
(with reheating as discussed). 

In Section \ref{pert} we analysed in detail the evolution of perturbations
in $\phi$CDM. We found that perturbations in the scalar field
on subhorizon scales decayed, leading to a suppression of power
on small scales. The similarities with MDM led us to pursue the
comparison in more detail. We found that the contribution from
the scalar field was more efficient at suppressing perturbations in
the CDM than massive neutrinos. We showed that this was due
to a simple difference in the evolution of the ``free-streaming'' scale
in the two theories: In $\phi$CDM the free streaming scale grows
with the horizon while in MDM the freestreaming scale decays as
$1/\tau$. This means that perturbations in the scalar field
{\it never} grow once they come into the horizon, in contrast 
to perturbations in the massive neutrino which end up clumping after
some finite time. We analysed the effect the scalar field
would have on the CMB and found it to be small but distinctly
different from that of MDM.

In Section \ref{data} we used the results of a Boltzmann-Einstein
solver to test how well this class of models fared when compared
to various astrophysical data. Because of the weak effect the 
scalar field has on the CMB, the angular power spectrum is 
effectively (with the current accuracy of experiments) as in SCDM. 
Using the COBE data we normalized these 
theories and compared the
mass variance per logarithmic interval in $k$ to the one  estimated
in \cite{PD}. Our models fared as well, or better, than competing
flat universe models. A comparison with an estimate of the
mass variance at $8$h$^{-1}$Mpc from the abundances of rich 
clusters gave the same results. We finally
compared the amount of structure at high redshift our model
predicts, as compared to that infered from Lyman-$\alpha$ systems.
This has proven to be a serious problem for MDM models. We
saw that our model is consistent, albeit marginally, with
these constraints.

Lastly a few further commments on other related issues which 
it would be interesting to investigate:

(i) We assumed an initial flat adiabatic spectrum of perturbations,
in line with the most generic type of inflation.  Within the
context of inflation one can of course have different spectra etc.,
and within the context of some very well motivated form for the 
inflationary part of the potential in the alternative reheating model,
it would be interesting to look at the combined effect on structure
formation. In more general,  an interesting feature of the exponential
which would be worth investigating is the fact that the attractor
is also an attractor for isocurvature fluctuations, and hence the
assumption of adiabatic initial conditions might have a much more
general motivation than the standard inflationary one.

(ii) Inflation was assumed simply because it is the
paradigmatic model. The existence of the exponential scalar field
might of course have an effect on any cosmology. In an open 
cosmology, for example, the analagous attractor also exists
in the curvature dominated regime and the asymptotic state
has a scalar field energy scaling as $1/a^2$. 

(iii) We have shown that the effect of the exponential
scalar field on the angular power spectrum of the CMB is quite
small, much like the case of MDM. However, recent high precision
analysis of parameter estimation from the CMB (as one would 
expect from the satellite missions) indicates that one may
achieve a precision of $\Delta \Omega_\nu$=0.04 or better.
This opens up the interesting possibility of actually trying
to constrain $\Omega_\phi$ with the CMB.

(iv)  Besides the issue of consistency with entry into the attractor
prior to nucleosynthesis which would motivate the study of the dependence
on $\Omega_bh^2$ away from the SCDM value we assumed, there are 
further observational reasons for doing so. The recent measurements 
of \cite{Tytler} indicate that one may have a higher baryon content 
than previously expected. Since a higher baryon content leads to less
structure on smaller scales, a best fit to large scale stucture constraints
would be obtained with a smaller value of $\Omega_\phi$. The effect on
the Lyman-$\alpha$ constraints might be more dramatic, since there would
be a competition between the suppression of power and the increase in the 
amount of gas simply due to the fact that there are
more baryons around. That the result is not immediately evident can be seen
from the analysis of MDM in \cite{he}. Further, to determine whether
entry to the attractor prior to nucleosynthesis is consistent in this 
case, one would have to determine whether this decrease in $\Omega_\phi$
would broaden the allowed range in $\eta$ sufficiently.

(v) One of the main problems for the MDM model 
is the overwhelming evidence for structure at high redshift.
Critical universe models with massive neutrinos typically
underproduce structure as probed by Lyman-$\alpha$ systems and
high redshift cluster abundances \cite{bahcall}. This problem is somewhat
alleviated in the case of $\phi$CDM: The different evolution
of the free-streaming scale in $\phi$CDM leads to more
power on small scales (and consequently high redshifts) relative
to MDM. This certainly provides strong motivation for
further study of the evolution of perturbations on small
scales and high redshift in these models.

From the point of view of structure formation we have described
a new model which has the same qualitative features as MDM. Unlike
other scalar field cosmologies, which affect the local expansion
rate (up to redshift of a few), our background evolution is
exactly matter dominated, the modification arising at the
perturbation level. Given the wealth of current data on small
scales, at recent redshifts, this raises the question of whether
this is the right approach to the construction models of
structure formation: The existence of structure at high redshift
combined with the small scale velocity dispersion today at
1h$^{-1}$Mpc seems to argue for a strong modification of
the growth rate of perturbations in the last few redshifts.
This would point towards a low density universe.
However, until we have a more detailed understanding of the non-linear
evolution of perturbations in models such as $\phi$CDM (on scales
between $0.1$ and $10$h$^{-1}$Mpc), these models should not be ruled
out. We are currently analysing the non-linear regime of $\phi$CDM
using an N-body code.

ACKNOWLEDGMENTS: We thank C. Balland,  
M. Davis,  J. Levin, A. Liddle, A. Jaffe, T. Prokopec, E. Scannapieco,
J. Silk and R. Taillet for useful discussions.
P.F. was supported by the
Center for Particle Astrophysics, a NSF Science and
Technology Center at U.C.~Berkeley, under Cooperative
Agreement No. AST 9120005. P.F thanks PRAXIS XXI (Portugal) and
CNRS (France) for partial support. MJ is supported by an
Irish Government (Dept. of Education) post-doctoral fellowship.

\pagebreak
\pagestyle{empty}


\begin{thebibliography}{99}
\bibitem{fail}Extrapolating the behaviour of these models
over such a large range of scales should be done with care.
Variations of $\Delta \Phi>M_P$ are beyond the range in which
the approximations we are considering are valid with any accuracy.
\bibitem{commBBKS}Note that the BBKS \cite{BBKS}
 fitting function is not accurate
with non-zero baryon content. For this, see \cite{Ma}.
\end{thebibliography}
\end{document}